\documentclass{article}
\usepackage[english]{babel}
\usepackage[T2A]{fontenc}

\usepackage[tbtags]{amsmath}
\usepackage{amsfonts,amssymb,mathrsfs,amscd}
\usepackage{tikz}
\usepackage{hyperref}

\newcommand\scalemath[2]{\scalebox{#1}{\mbox{\ensuremath{\displaystyle #2}}}}

\numberwithin{equation}{section}
\newtheorem{theorem}{Theorem}

\newtheorem{propos}{Proposition}
\newtheorem{definition}{Definition}

\def\g{\color{orange}}

\date{}

\begin{document}
    \title{Canonical forms of metric graph eikonal algebra and graph geometry}
    \author{M.I. Belishev\footnote{PDMI RAS,\,\,{belishev@pdmi.ras.ru}}, A.V. Kaplun\footnote{PDMI RAS,\,\,alex.v.kaplun@gmail.com,\,\,kaplunav@pdmi.ras.ru}}

    \maketitle
\begin{abstract}
The algebra of eikonals $\mathfrak E$ of a metric graph $\Omega$
is an operator $C^*$-algebra determined by dynamical system with
boundary control that describes wave pro\-pa\-ga\-tion on the
graph. In this paper, two canonical block forms ({\it algebraic}
and {\it geometric}) of the algebra $\mathfrak E$ are provided for
an arbitrary connected locally compact graph. These forms
determine some metric graphs ({\it frames}) $\mathfrak F^{\,\rm
a}$ and $\mathfrak F^{\,\rm g}$. Frame $\mathfrak F^{\,\rm a}$ is
determined by the boundary inverse data. Frame $\mathfrak F^{\,\rm
g}$ is related to graph geometry. A class of {\it ordinary graphs}
is introduced, whose frames are identical: $\mathfrak F^{\,\rm
a}\equiv\mathfrak F^{\,\rm g}$. The results are supposed to be
used in the inverse problem that consists in determination of the
graph from its boundary inverse data.
\end{abstract}

\footnotetext{This work was supported by the Leonhard Euler
Mathematical Institute, agreement no. 075-15-2022-289, grant RFBR
18-01-00269, and partially by the Young Mathematicians of Russia
contest.}

    \section{Introduction}\label{Introduction}
\noindent$\bullet\,\,\,$ The eikonal algebra of the metric graph
was introduced in \cite{Belishev_Wada_2015} and studied in
\cite{Belishev_Kaplun_2018,Belishev_Kaplun_2022,Kaplun_2021,Kaplun_dissert}.
It is a noncommutative operator $C^*$-algebra determined by adynamical system that describes the propagation of waves into the
graph; the waves are initiated by sources (controls) at the boundary vertices. The interest to such an algebra is motivated by
possible applications to inverse problems, in particular for graph
reconstruction from boundary spectral and dynamical inverse data.
The relation between inverse problems and Banach algebras is a
separate topic within the {\it boundary control method}
(BC-method; \cite{Bel_UMN}).
\smallskip

\noindent$\bullet\,\,\,$ In our paper, for an arbitrary connected
locally compact graph $\Omega$ and its corresponding eikonal
algebra $\mathfrak{E}_{\Sigma}$, we describe two of its canonical
block forms --- {\it algebraic} and {\it geometric}. Both forms
are derived from the original parametric representation of the
algebra $\mathfrak{E}_{\Sigma}$. The algebraic form is known and
has been described in details in \cite{Belishev_Kaplun_2022} (see
also examples in \cite{Belishev_Kaplun_2018,Kaplun_2021}). The
geometric form is a new one.
\smallskip

\noindent$\bullet\,\,\,$ Our results are as follows.

It is shown that both forms determine some metric graphs{\g:} {\it
frames} ${\mathfrak F}^{\,\rm a}_\Sigma$ and ${\mathfrak F}^{\,\rm
g}_\Sigma$ respectively.

The frame ${\mathfrak F}^{\,\rm a}_\Sigma$ is the spectrum (the
set set of irreducible representations) of $\mathfrak{E}_{\Sigma}$
factorized by some relation and equipped with relevant
coordinates. It is an invariant of the algebra
$\mathfrak{E}_{\Sigma}$: any of its isomorphic copy determines
frame ${\mathfrak F}^{\,\rm a}_\Sigma$ up to isometry of metric
spaces.

The frame ${\mathfrak F}^{\,\rm g}_\Sigma$ is directly related to
the graph geometry; more precisely, to the shape of the domain of
the graph filled with waves. It is the result of factorization
(gluing) of this domain according to a relation which has a
transparent geometric meaning. Frame ${\mathfrak F}^{\,\rm
g}_\Sigma$ also gets relevant coordinates which have the
same nature as in the case of the frame ${\mathfrak F}^{\,\rm
a}_\Sigma$.

The concept of {\it ordinary graphs} is introduced. Their frames
${\mathfrak F}^{\,\rm a}_\Sigma$ and ${\mathfrak F}^{\,\rm
g}_\Sigma$ are identical: the bijection linking points of frames
with the same coordinates turns out to be an isometry.
\smallskip

\noindent$\bullet\,\,\,$ Possible applications of our results to
inverse problems are discussed in Comments at the end of the
paper. So far there are no successful applications, that explains
the lack of references to the extensive literature on inverse
problems on graphs.

The introductory part of the paper has significant overlaps with
the materials of previous papers on the eikonal algebra. This is
justified by the wish to make the paper as independent as
possible.

    \section{Waves on graph}
    \subsubsection*{Metric graph}

\noindent$\bullet\,\,\,$ Let $I_j:=
(0,a_j)=\{s_j\in\mathbb{R}\,|\,\, 0<s_j<a_j<\infty\}, j=1,\dots,d$
be finite intervals. The set  $S_d:=\{0\}\sqcup
I_1\sqcup\dots\sqcup~I_d$ equipped with the metric
    \begin{equation}\label{metric tau}
        \tau(x,y):=\left\{\begin{array}{ll}
            |x-y|, & \text{in }x,y\in I_j\\\
            x+y, & \text{at }x\in I_i,\,\, y\in I_j, i\neq j\\\
            y, & \text{at }x=0,\,\,y\in I_j\\\
            x, & \text{at }x\in I_i, \,\,\,y=0\\
            0, & \text{in }x=y=0
        \end{array}\right.
    \end{equation}
is said to be a \textit{d-star}.
\smallskip

\noindent$\bullet\,\,\,$ \textit{Metric graph} $\Omega$ is a
connected metric space which is locally isometric to either a star
or an interval. Its interior vertices are points whose (small)
neighborhoods are isometric to stars $S_d$ with $d>2$; the
boundary vertices correspond to stars $S_1$ (semi-open intervals).
The number $d$ is called the \textit{valence} of a vertex. The
absence of  vertices with valence $2$ in the further
considerations is explained by the fact that 2-star is isometric
to an interval. Therefore the part of the graph that is isometric to $S_2$ can be replaced by the appropriate interval. However, below, while considering frames, the vertices with $d=2$ are used. The edges are the maximal parts of $\Omega$ that
do not contain vertices and are isometric to intervals.

Thus, $\Omega=E\sqcup V\sqcup \Gamma$, where $E=\{e_i\}_{i=1}^p$
are \textit{edges}; $V=\{v_j\}_{j=1}^q$, $v_k$ -- \textit{internal
vertices}; $\Gamma=\{\gamma_k\}_{k=1}^{n}$, $\gamma_k$ --
\textit{boundary vertices}; each vertex $w \in V\sqcup \Gamma$ has
a neighborhood in $\Omega$ that is isometric to $S_d$. Further, we
assume that the set of boundary vertices is not empty.

The given definition admits the edges of infinite length.
\smallskip

\noindent$\bullet\,\,\,$ Let $\tau$ be a metric on $\Omega$,
$A\subset\Omega$ a subset. By
$\Omega^r[A]:=\{x\in\Omega\,|\,\,\,\tau(x,A)<r\}$ we denote the
metric neighborhood of radius $r>0$ of $A$.

\subsubsection*{Operators and spaces on graph}
\noindent$\bullet\,\,\,$ Let us orient each of the edges by a
linear order $\prec$ (in one of two possible ways). For the edge
$e\in E$, the function $y$ on $\Omega$ and the point $x\in e$, we
define the derivative
    $$
    \frac{dy}{de}(x)\,:=\,\lim\limits_{m\to x}
    \frac{y(m)-y(x)}{s_m\tau(m,x)}\,,
    $$
where $s_m=1$ for $x\prec m$ and $s_m=-1$ for $m\prec x$. { Note also that the value of the second derivative
$\frac{d^2y}{de^2}$ is independent from the orientation of the
edges.}

Let us choose a vertex $w\in V\cup \Gamma$ and its neighborhood in
$\Omega$, which is isometric to a star $S_d$. We say that an edge
$e$ is adjacent to $w$ if $w\in \overline{e}$ \,(closure in
$\Omega$). For each $e$ adjacent to $w$, we define
\textit{outgoing derivative}
    \begin{equation*}
        \frac{dy}{de_+}(w):=\lim_{e\ni m\to
            w}\frac{y(m)-y(w)}{\tau(m,w)},
    \end{equation*}
which does not depend on the orientation. For each vertex $w\in
V\cup \Gamma$ and function $y$ we define \textit{outgoing  flow}
    \begin{equation*}
        \Pi_{w}[y] = \sum_{\overline{e}\ni w}   \frac{dy}{de_+}(w).
    \end{equation*}
\smallskip

\noindent$\bullet\,\,\,$ Consider a real Hilbert space
$\mathscr{H}:=L_2(\Omega)$ of functions on $\Omega$ with the
scalar product
    \begin{equation*}
        (y,u)_{\mathscr{H}}=\int_{\Omega} yu\,
        d\tau=\sum_{e\in E}\int_{e} yu \,d\tau.
    \end{equation*}
By $C(\Omega)$ we denote the space of continuous functions with
the norm $\|y\|= \sup\limits_\Omega|y(\cdot)|$. We assign the
function $y$ to the Sobolev class $\mathscr{H}^2(\Omega)$ if $y\in
C(\Omega)$ and $\frac{dy}{de},\frac{d^2y}{de^2}\in L_2(e)$ on each
edge $e\in E$.

Introduce the Kirchhoff class
    \begin{equation*}
        {\mathscr K} \,:= \,\{y \in {\mathscr H}^2(\Omega)\,|\,\,\,\Pi_v[y]=0,\,\,\,v
        \in V \}\,.
    \end{equation*}

The Laplace operator on the graph is introduced via the definition
    \begin{equation}\label{Eq Laplace operator}
        \Delta: {\mathscr  H}\to{\mathscr H}; \quad \textrm{ Dom\,}
        \Delta={\mathscr K};\quad\left(\Delta y\right
        )\big|_e=\frac{d^2y}{de^2}\,, \,\,\, e \in E.
    \end{equation}
It is densely defined, closed, and does not depend on
the orientation of the edges.

\subsubsection*{Dynamical system with boundary control}

\noindent$\bullet\,\,\,$ The boundary value problem that describes
wave propagation in the graph, is of the form
    \begin{align}
        \label{Eq DS 1} & u_{tt}-\Delta u=0 && \text{in}\,\,\,{\mathscr H},
        \,\,\,0<t<T;\\
        \label{Eq DS 2} &u(\cdot,t)\in {\mathscr K}  && \text{for}\,\,\,\,0\leqslant t\leqslant T;\\
        \label{Eq DS 3} & u|_{t=0}=u_t|_{t=0}=0 && \text{in}\,\,\, \Omega;\\
        \label{Eq DS 4} & u=f && \text{on}\,\,\,\, \Gamma \times [0,T].
    \end{align}
Here $T >0$ is end time;\, $f=f(\gamma ,t)$ is \textit{boundary
control}; $u=u^f(x,t)$ is the solution (\textit{wave}). With
$C^2$-smooth (with respect to $t$) control $f$ disappearing near
$t=0$, the problem has a unique classical solution $u^f$.

As follows from the definition (\ref{Eq Laplace operator}), on
each edge $e$ the solution $u^f$ satisfies the homogeneous string
equation $u_{tt}-u_{ee}=0$. Therefore, the waves propagate from
the boundary $\Gamma$ into $\Omega$ with unit velocity. As a
consequence, if the control operates from a part of the boundary
$\Sigma\subseteq\Gamma$, i.e.,
$\textrm{supp\,}f\subset\Sigma\times[0,T]$ is satisfied then we
have the relation
 \begin{equation}\label{Eq supp u^f(cdot,t) Sigma}
        \textrm{ supp\,}u^f(\cdot,t) \subset \overline{\Omega^t[\Sigma]},
        \qquad t>0\,.
    \end{equation}

\noindent$\bullet\,\,\,$ The control space
$\mathscr{F}^T:=L_2(\Gamma\times  [0,T])$ with the scalar product
\begin{equation*}
(f,g)_{\mathscr{F}^T}:=\sum_{\gamma\in\Gamma}\int_0^T
f(\gamma,t)\, g(\gamma,t)\, dt
\end{equation*}
is called an \textit{external space} of the system (\ref{Eq DS
1})-(\ref{Eq DS 4}). It is represented in the form
\begin{equation*}
\mathscr{F}^T=\oplus\sum_{\gamma\in\Gamma}\mathscr{F}^T_{\gamma}
\end{equation*}
as a sum of subspaces
$\mathscr{F}^T_{\gamma}:=\{f\in\mathscr{F}^T|\,\textrm{
supp}\subset\{\gamma\times[0,T]\}\}$.

The space $\mathscr{H}=L_2(\Omega)$ is called \textit{internal},
the waves $u^f(\cdot,t)$ are its time-dependent elements.

\subsubsection*{Eikonals} Let us choose the boundary vertex
$\gamma\in\Sigma$ and fix a final moment $t=T$. The set of waves
\begin{equation*}
{\mathscr U}^s_\gamma:=\left\{u^f(\cdot,s)\,|\,\,f \in {\mathscr
F}^T_\gamma\right\}\,\subset\,{\mathscr H}, \qquad 0\leqslant
s\leqslant T
\end{equation*}
is called \textit{reachable} (from vertex $\gamma$, at time
$t=s$). It can be shown that ${\mathscr U}^s_\gamma$ are (closed)
subspaces in $\mathscr H$. As $s$ grows, they expand: ${\mathscr
U}^s_\gamma\subset{\mathscr U}^{s'}_\gamma$ at $s<s'$. Let
$P^s_{\gamma}$ in $\mathscr{H}$ be (orthogonal) projectors on
subspaces $\mathscr{U}^s_\gamma$. The family
$\{P^s_\gamma\,|\,\,0\leqslant s\leqslant T\}$ is continuous with
respect to $s$ and defines an eikonal operator (briefly --
\textit{eikonal})
    \begin{equation*}
        E_\gamma:\,\mathscr{H}\rightarrow \mathscr{H},\quad E_\gamma\,:=\,\int^{T}_0(s+1)\,dP^s_\gamma.
    \end{equation*}
\cite{Belishev_Wada_2015}. From the definition it follows that
$E_\gamma$ is a bounded self-adjoint positive operator. Let {
$T^*_{\gamma}:=\sup\limits_{\gamma'\in\Gamma}\tau(\gamma,\gamma')\leqslant
\infty$} be the time of filling the graph with waves from the
vertex $\gamma$ (see (\ref{Eq supp u^f(cdot,t) Sigma}). The
following statement appears to be true
 \begin{propos}
Eikonal $ E_{\gamma}$ satisfies ${\rm Ran\,} E_{\gamma}
=\mathscr{U}^{T}_\gamma$ and ${\rm Ker\,} E_{\gamma} =
\mathscr{H}\ominus\mathscr{U}^{T}_\gamma$. At $T<T^*_\gamma$ it
has eigenvalue $0$ of infinite multiplicity and a simple
absolutely continuous spectrum filling the segment $[1,T+1]$.
    \end{propos}

Note that existence of at least one edge of infinite length leads
to the fact that the spectrum of the eikonal has no lacunas and
for any $T<\infty$ the equality
\begin{align}
\notag & \sigma(
E_\gamma\big|_{\mathscr{U}^{T}_\gamma})=\sigma_\textrm{ac}(
E_\gamma)=[1,T+1]
\end{align}
holds \cite{Belishev_Kaplun_2022}. {The existence of lacunae
in the spectrum in the case $T^*_\gamma<T<\infty$ for compact graphs is an open
question.}

\section{Algebra of eikonals}\label{sec2}

\subsubsection*{On algebras}

The following information about $C^*$-algebras is taken from
\cite{Arveson,Naimark,Dixmier,Murphy}.
\smallskip

\noindent$\bullet\,\,\,$ Recall that a Banach algebra $\mathfrak
A$ with involution $(\cdot)^*$ is called a $C^*$-algebra if
    \begin{equation*}
        \|a^* a\|\, = \,\|a\|^2,\qquad a\in\mathfrak A
    \end{equation*}
holds. In particular, such are the algebras of bounded operators
$\mathfrak{B}(H)$ in the Hilbert space $H$, in which the role of
involution plays the operator conjugation. All algebras in the
paper are operator $C^*$-algebras \cite{Murphy}.

The notation ${\mathfrak A}\cong{\mathfrak B}$ will mean that
$C^*$-algebras ${\mathfrak A}$ and ${\mathfrak B}$ are related by
$*$ - isomorphism ({hereafter briefly -- \textit{isomorphism}}).
Such isomorphism is an isometry.

For the set $S\subset \mathfrak{A}$, by $\vee S$ we denote the
minimal $C^*$-(sub)algebra in $\mathfrak{A}$ that contains $S$.
\smallskip

\noindent$\bullet$\,\,\, For the algebra $A=A_1\oplus\dots\oplus
A_p$ , by ${\rm pr}_j$ we denote the projections {
$a_1\oplus\dots\oplus a_j\oplus\dots a_p\mapsto a_j$}. Let
$C\subset A$ be a subalgebra; the algebras ${\rm pr}_j\,C$ are
called the {\it blocks} of $C$.

We say that algebra $C$ {\it separates} the blocks
$A_{i_1},\dots,A_{i_m}$ of algebra $A$ if, for any set of elements
{$a_{i_1}\in A_{i_1},\dots,a_{i_m}\in A_{i_m}$,} one can find
the elements $c_{i_1},\dots,c_{i_m}$ in $C$ that satisfy
$$
{\rm pr}_{i_l}\,c_{i_k}= \begin{cases}
                a_{i_k}, & i_k=i_l;\cr
                0_{i_l}, & i_k\not=i_l;\cr
                         \end{cases}\,\,.
$$
If algebra $A$ has the blocks that are not separated by algebra
$C$, we say that $C$ {\it links} blocks of $A$.
\smallskip

\noindent$\bullet\,\,\,$ A representation of $C^*$-algebra
$\mathfrak{A}$ is a $*$-homomorphism
$\pi:\mathfrak{A}\to\mathfrak{B}(H)$. The equivalence of the
representations $\pi\sim\pi'$ means that
$\iota\pi(a)=\pi'(a)\iota$, $a\in\mathfrak{A}$, where $\iota: H\to
H'$ is a unitary operator. The representation is
\textit{irreducible} if the operators $\pi(\mathfrak{A})$ share no
nonzero invariant subspace in $H$.

The spectrum of $C^*$-algebra $\mathfrak{A}$ is the set
$\widehat{\mathfrak{A}}$ of equivalence classes of its irreducible
representations. The equivalence class (a point of the spectrum)
corresponding to the representation $\pi$ will be denoted by
$\hat\pi$. The spectrum is equipped with the canonical Jacobson
topology \cite{Dixmier, Murphy}.

Isomorphism of algebras $\textrm{ u}:\mathfrak{A}\to\mathfrak{B}$
induces the correspondence between the representations
    \begin{align*}
       { \pi\mapsto\textrm{ u}_*\pi,\quad(\mathrm{u}_*\pi)(b):=\pi(\textrm{u}^{-1}(b)),}
    \end{align*}
which extends to the canonical homeomorphism of the spectra:
    \begin{align*}
        \widehat{\mathfrak{A}}\ni \hat\pi\, {\mapsto} \textrm{ u}_*\hat\pi\in\widehat{\mathfrak{B}},\quad
        \textrm{ u}_*\hat\pi:=\{\textrm{ u}_*\pi\,|\,\,\pi\in\hat\pi\}.
    \end{align*}

\subsubsection*{Standard algebras and their spectra}
\noindent$\bullet\,\,\,$ By $\mathbb{M}^n$ we denote the algebra
of the real $n\!\times\! n$-matrices viewed as operators in
$\mathbb R^n$ and equipped with the corresponding (operator) norm.
This algebra is irreducible.

The $C^*$-subalgebra $\mathfrak A\subset \mathbb{M}^n$ can also be
considered irreducible if $\mathfrak A\cong \mathbb{M}^k$, where
$k\leqslant n$ holds. Such an algebra, in a suitable basis in
$\mathbb R^n$, takes a block-diagonal form and is composed of two
blocks, one of which is $\mathbb{M}^k$ and the second (if exists)
is zero.
    \begin{propos}\label{Prop on M^n}
Any $C^*$-subdalgebra of the algebra $\mathbb{M}^n$ is isomorphic
to a direct sum $\bigoplus_{k}\mathbb{M}^{n_k}$, where $\sum_{k}
{n_k}\leqslant n$.
    \end{propos}
    \smallskip

\noindent$\bullet\,\,\,$ By $C([a,b],\mathbb{M}^n)$ we denote the
algebra  of continuous $\mathbb{M}^n$-valued functions with the
norm $\|\phi\|=\underset{a\leqslant t\leqslant  b}
{\textrm{sup\,}}\|\phi(t)\|_{\mathbb{M}^n}$. By the same symbol we
denote the (sub)algebra in $\mathfrak{B}\left(L_2([a,b];\mathbb
R^n)\right)$ of the operators that multiply square-integrable
$\mathbb R^n$-valued functions by functions from
$C([a,b],\mathbb{M}^n)$. The correspondence $\phi\mapsto
\phi\,\cdot$\,\, establishes the isomorphism of these algebras.

The following fact plays an important role (see, e.g.,
\cite{Murphy}).
    \begin{propos}\label{Prop p as px} The representations
        \begin{equation}\label{Eq hom pi t}
            \pi_t:C([a,b],\mathbb{M}^n)\to\mathbb{M}^n;\,\,\,\pi_t(\phi):=\phi(t),\quad a\le t\le b
        \end{equation}
are irreducible; their equivalence classes exhaust the spectrum of
the algebra $C([a,b],\mathbb{M}^n)$. For any irreducible
representation $\pi$ of the algebra $C([a,b],\mathbb{M}^n)$ there
exists a single point $t\in[a,b]$ such that $\pi\sim\pi_t$.
    \end{propos}

Algebra $C([a,b]; \mathbb M^n)$ contains subalgebras
    \begin{equation}\label{standard alg}
        \dot C([a,b]; \mathbb M^n):=\left\{\phi\in C([a,b]; \mathbb
        M^n)\,|\,\,\,\phi(a)\in \mathbb M_a,\,\,\,\phi(b)\in \mathbb
        M_b\right\},,
    \end{equation}
where $\mathbb M_a,\mathbb M_b$ are $C^*-$subalgebras of $\mathbb
M^n$, which we call \textit{boundary algebras}. Proposition
\ref{Prop on M^n} implies
        \begin{equation}\label{Eq boundary algebras standard}
            \mathbb M_a\cong\bigoplus\limits_{k=1}^{n_a} \mathbb
            M^{\kappa_k},\,\,\,\kappa_1+\dots+\kappa_{n_a}\leqslant
            n; \quad \mathbb M_b\cong\bigoplus\limits_{k=1}^{n_b}\mathbb
            M^{\lambda_k},\,\,\,\lambda_1+\dots+\lambda_{n_b}\leqslant n.
        \end{equation}
In the case $\mathbb M_a=\mathbb M_b=\mathbb M^n$ we have $\dot
C([a,b]; \mathbb M^n)= C([a,b]; \mathbb M^n)$. We call $\dot
C([a,b]; \mathbb M^n)$ \textit{standard algebras}.
    \smallskip

\noindent$\bullet\,\,\,$ The standard algebra spectrum consists of
the classes $\hat\pi_t,\,\,t\in(a,b)$ of irreducible
representations of the form (\ref{Eq hom pi t}), and {the
calsses $\hat\pi^k_a,\,\hat\pi^k_b$, into which the  representations of the boundary algebras
$\mathbb{M}_a,\mathbb{M}_b$ (generally
speaking, reducible) are decomposed.} If, for example,
$n_a\geqslant 2$, then $\pi_a$ decomposes into irreducible
representations
    \begin{equation}\label{Eq boundary hom}
        \pi_a^k:\phi(a)\mapsto [\phi(a)]^k\in\mathbb M^{\kappa_k},
    \end{equation}
where $[...]^k$ is the $k$-th block of the block-diagonal matrix
in representations (\ref{Eq boundary algebras standard}). In this
case we say that the representations
$\hat\pi_a^1,\dots,\hat\pi_a^{n_a}$ form a \textit{cluster} in the
spectrum of the standard algebra. This term is motivated by the
fact that they are inseparable from each other in the Jacobson
topology \cite{Arveson}. A similar cluster may exist at the
right-hand end at $t=b$. At the same time, all $\hat\pi_t$ with
different $t\in(a,b)$ are separable from each other and from
clusters (see \cite{Belishev_Wada_2015},
\cite{Belishev_Kaplun_2018}). The spectrum of the algebra
$C([a,b]; \mathbb M^n)$ contains no clusters.

The preceding facts and results are taken from \cite{Arveson,Naimark,Dixmier,Murphy}.
    \smallskip

\noindent$\bullet\,\,\,$ Here is a general concept. Let $\sim_0$
be a symmetric reflexive {(but, may be, not transitive)
relation on a set $X$. Consider on $X$ the relation $\sim$ that is
defined by the following rule: $x\sim y$ if there is a finite set
of elements $x_1,\dots,x_n$ such that $x\sim_0 x_1\sim_0
\dots\sim_0 x_n\sim_0 y$ holds. This relation will be called the
\textit{transitive closure} of the relation $\sim_0$. One can
easily see that $\sim$ is an equivalence relation.}

\subsubsection*{Algebra $\mathfrak{E}_{\Sigma}$}
In the rest of the paper, unless otherwise specified, the final
moment $T$ in (\ref{Eq DS 1})--(\ref{Eq DS 4}) is considered fixed
and is not always indicated in the notation.
\smallskip

\noindent$\bullet\,\,\,$ Let us choose a subset
$\Sigma=\{\gamma_1,\dots,\gamma_N\}\subset\Gamma$; the vertices,
which it consists of, are called {\it controlling}. Due to
(\ref{Eq supp u^f(cdot,t) Sigma}) we have
    \begin{equation*}
        {\mathscr U}^s_\gamma\,\subset\,{\mathscr H}^T_\Sigma\,:=\{{ y\in\mathscr H}\,|\,\,{\rm supp\,}y\subset
        \overline{\Omega^T[\Sigma]}\}, \qquad \gamma\in\Sigma,\,\,\,0\leqslant
        s\leqslant T.
    \end{equation*}
    As a consequence, for a full reachable set we have the embedding
    \begin{equation*}
        {\mathscr U}^s_\Sigma\,:=\,\overline{{\rm span\,}\{{\mathscr
            U}^s_\gamma\,|\,\,\gamma\in\Sigma\}}\,\subset\,{\mathscr
            H}^T_\Sigma,\qquad 0\leqslant s\leqslant T.
    \end{equation*}
Thus, if the controls $f$ act from the vertices $\gamma\in\Sigma$
only, then the natural interior space of the system (\ref{Eq DS
1})--(\ref{Eq DS 4}) is the subspace {${\mathscr
H}^T_\Sigma=L_2({\Omega^T[\Sigma]})$}. Eikonals $E_\gamma$, which
correspond to the vertices of $\gamma\in\Sigma$, are the operators
in ${\mathscr H}^T_\Sigma$.
\smallskip

\noindent$\bullet\,\,\,$ By the eikonal algebra we call the
operator $C^*$-algebra
\begin{equation*}
\mathfrak{E}_{\Sigma}:=\vee\{E_{\gamma}\,|\,\,\gamma\in\Sigma\}\subset\mathfrak{B}({\mathscr
H}^T_\Sigma),
\end{equation*}
introduced in \cite{Belishev_Wada_2015}. The new results relating
to it, which constitute the main subject of our paper, are placed
in section \ref{sec3}. Their presentation requires certain
preparation, which is done in the remaining part of the section
\ref{sec2} and in the section \ref{sec*}, in which known facts and
results are briefly described. The reader can find more details in
\cite{Belishev_Wada_2015,Belishev_Kaplun_2022}.

\subsubsection*{Graph parametrization}
\noindent$\bullet\,\,\,$ A set
$\Lambda=\{x_1,\dots,x_m\}\subset{\overline{\Omega^T[\Sigma]}}$ is
called {\it determination set}, if for any function $y\in
C(\overline{\Omega^T[\Sigma]})$ and for all $\gamma\in\Sigma$, the
values of functions $E_\gamma y$ on $\Lambda$ are determined by
the values of $y$ on $\Lambda$. Equivalently, the implication
takes place
    \begin{equation}\label{det set}
        y|_\Lambda=0\,\Rightarrow\,(E_\gamma y)|_\Lambda=0,\qquad
        \gamma\in\Sigma.
    \end{equation}
Every point ${x\in}\overline{\Omega^T[\Sigma]}$, with the
possible exception of {\it a finite} number of so-called critical
points, belongs to some (generally speaking, not unique)
determination set $\Lambda[x]$: see
\,\cite{Belishev_Wada_2015,Belishev_Kaplun_2022}.
\smallskip

\noindent$\bullet\,\,\,$ As it follows from (\ref{det set}), the
pairs $\{y|_\Lambda,(E_\gamma y)|_\Lambda\}$ constitute the graph
of an operator $e_\gamma$, which acts in the $m$-dimensional space
${\bf l}_2(\Lambda)$ of functions (vectors) with the scalar
product $( a, b)=\sum\limits_{k=1}^m a(x_k)\,b(x_k)$. The
correspondence $\pi: E_\gamma\mapsto e_\gamma$ extends from the
generators $E_\gamma,\,\,\gamma\in\Sigma$\, to the whole algebra
$\mathfrak{E}_{\Sigma}$ and provides its $m$-dimensional
representation
    \begin{equation}\label{repr 1}
        \pi:\,\mathfrak{E}_{\Sigma}\,\to\,\mathfrak B({\bf l}_2(\Lambda)).
    \end{equation}
Let $\{\chi_1,\dots,\chi_m\},\,\,\,\chi_i(x_k)=\delta_{ik}$ be the
basis in ${\bf l}_2(\Lambda)$ of indicators of the points forming
$\Lambda$. In this basis, operators $e_\gamma$ get the matrices
$\check e_\gamma$. Consequently, the representation (\ref{repr 1})
takes a matrix form:
    \begin{equation}\label{repr 2}
        \pi:\,\mathfrak{E}_{\Sigma}\,\to\,\mathbb M^{\,m}.
    \end{equation}

As will be seen later, almost all representations of the eikonal
algebra are of the form (\ref{repr 1}), (\ref{repr 2}).
\smallskip

\noindent$\bullet\,\,\,$ When the point $x\in\Lambda[x]$ is
smoothly varied, the points $x_1,\dots,x_m$ composing
$\Lambda[x]$, vary and cover open intervals ({\it cells})
$\omega_1,\dots,\omega_m$ of the equal length
$|\omega_k|=\epsilon>0$, which are located on the graph edges. The
cells form a {\it family}
    $$
\Phi:=\bigcup\limits_{k=1}^m\omega_k=\bigcup\limits_{x\in\omega}\Lambda[x],
    $$
where $\omega$ is any cell of the family $\Phi$.

The entire (sub)graph $\overline{\Omega^T[\Sigma]}$ that is filled
with waves is partitioned into a finite number of families
$\Phi^1,\dots,\Phi^J$ of this form
    \begin{equation}\label{partition}
        \Phi^j:=\bigcup\limits_{k=1}^{m_j}\omega^j_k,\qquad
        |\omega_k^j|=\epsilon_j
    \end{equation}
and a finite set $\Theta$ of the so-called {\it critical points}
that cover cell junctions:
    \begin{equation*}
        \overline{\Omega^T[\Sigma]}\,=\,\left[ \bigcup_{j=1}^{J}
        \Phi^j\right]\cup\Theta.
    \end{equation*}
As a consequence, there occurs the decomposition
    \begin{equation}\label{decomp L_2}
        L_2({\Omega^T[\Sigma]})\,=\,\oplus\sum\limits_{j=1}^J L_2(\Phi^j).
    \end{equation}

\noindent$\bullet\,\,\,$ Each family $\Phi^j$ in (\ref{partition})
is parameterized as follows. We take a cell $\omega_k^j\subset
\Phi^j$ and fix one of its endpoints $c\in\overline{\omega_k^j}$.
The point $x\in {\omega_k^j}$ gets the parameter
$r:=\tau(x,c),\,\,0< r<|\omega_k^j|$ and is denoted by $x_k(r)$:
    \begin{equation*}
        x_k(r)\,:=\,\Lambda[x(r)]\,\cap\,\omega^j_k,\qquad 0<
        r< \epsilon_j\,.
    \end{equation*}
As $r$ changes, the points $x_1(r),\dots,x_{m_j}(r)$ that
constitute the set $\Lambda[x(r)]$ move coherently along the edges
at unit speed and cover the corresponding (parameterized) cells
$\omega_k^j$.

The second of two possible parametrizations corresponds to the
choice of the other endpoint of the cell $\omega_k^j\subset
\Phi^j$. Other $\Phi^j$ are parametrized independently (in one of
two possible ways). In what follows, we assume that all families
$\Phi^j$ are parametrized.
    \smallskip

\noindent$\bullet\,\,\,$ These facts have a quite transparent
geometrical meaning associated with the so-called {\it hydra}
$H^T_\Sigma$ -- a space-time graph defined by the dynamical system
(\ref{Eq DS 1})--(\ref{Eq DS 4}): see \cite{Belishev_Wada_2015}
for details and illustrations.

Note in addition that the dynamics-related partition of the form
(\ref{partition}) is not unique: others can be obtained from it,
for example, by partitioning the families $\Phi^j$ into families
with smaller cells. The most interesting partition is the one that
corresponds to {\it minimal} (by $m=\#\Lambda$) sets $\Lambda$. The relations between minimality and irreducibility of representations (\ref{repr 1})
are discussed below.
\smallskip

\noindent$\bullet\,\,\,$ The parametrization of the graph induces
the parametrization of representations (\ref{repr 1}): we have the
series
    \begin{equation}\label{repr r}
        \pi_j(r):\mathfrak E_\Sigma\to\mathfrak B({\bf
            l}_2(\Lambda[x(r)])
        \quad x(r)\in
        \Phi^j,\,\,\,0< r<\epsilon_j;\,\,\, j=1,\dots,J.
    \end{equation}
It will be revealed later that representations of this type
essentially cover the spectrum of the algebra
$\mathfrak{E}_{\Sigma}$.

Further, along with (\ref{repr r}), it is convenient to use
concrete matrix-valued representations of the form (\ref{repr 2}).
For this purpose let us identify the spaces ${\bf
l}_2(\Lambda[x(r)])\equiv{\bf l}_2(\Lambda[x(r')]=: {\bf
l}^j_2\cong\mathbb R^{\,m_j}$ for $r,r'\in(0,\epsilon_j)$, the
indicator bases in them, and define the appropriate basis
$\{\chi_1,\dots,\chi_{m_j}\}$ in ${\bf l}^j_2$.

\subsubsection*{Parametrization of eikonal algebra}

Parametrization of the graph determines the parametric form of the
algebra $\mathfrak{E}_{\Sigma}$. Let us describe it according to
\cite{Belishev_Wada_2015}.
    \smallskip

\noindent$\bullet\,\,\,$ The partition (\ref{partition})
corresponds to the set of unitary operators $U^j: L_2(\Phi^j) \to
L_2((0,\epsilon_j),{\bf l}^j_2)$,
    \begin{equation}\label{unit U^j}
        (U^j y)(r)\,:=\,y\big|_{\Lambda[x(r)]}=
        \begin{pmatrix}
            y(x_1(r))\\\dots\\y(x_{m_j}(r))
        \end{pmatrix} \,\, \in\,{\bf
            l}^j_2,\quad r\in(0,\epsilon_j);\,\,\, j=1,\dots,J,
    \end{equation}
that define the unitary operator
    \begin{equation*}
        {U}: L_2(\overline{\Omega^T[\Sigma]}) \to \oplus\sum_{j=1}^J L_2((0,\epsilon_j);{\bf l}^j_2),
        \quad {U}:=\oplus\sum_{j=1}^J U^j.
    \end{equation*}
All eikonals $E_{\gamma}$ are reduced by subspaces $L_2(\Phi^j)$
from the decomposition (\ref{decomp L_2}), which expectedly
follows from $E_\gamma$-invariance of determination sets: see
(\ref{det set}). They turn into operators that multiply the
elements of the space of parametric representation
$\oplus\sum_{j=1}^J L_2((0,\epsilon_j);{\bf l}^j_2)$ by the
operator-valued functions
    \begin{equation}\label{UEU-1 first}
        {U}E_{\gamma}{U}^{-1} = \oplus\sum_{j=1}^J \left[\sum\limits_{i=1}^{n_{\gamma j}}
        \tau_{\gamma j}^i(\cdot_j) P_{\gamma j}^i\right] \in \bigoplus_{j=1}^J
        C\left([0,\epsilon_j];\mathfrak B({\bf
            l}^j_2)\right),\quad \mathfrak B({\bf
            l}^j_2)\cong \mathbb{M}^{m_j},
    \end{equation}
where each (scalar) function $\tau_{\gamma j}^i$ depends on its
argument $r_j\in[0,\epsilon_j]$ and is one of the following forms:
    \begin{equation}\label{functions tau}
        \begin{array}{l}
            \tau^i_{\gamma j}(r)=t^i_{\gamma j}\textbf{}+r \quad {\text{or}}\quad
            \tau^i_{\gamma j}(r)=\tilde{t}^i_{\gamma j}-r=(t^i_{\gamma
                j}+\epsilon_j)-r
        \end{array}
    \end{equation}
with the constants $t^i_{\gamma j},\tilde{t}^i_{\gamma j}$:
    \begin{equation*}
        t^i_{\gamma j}:=\min_{r\in [0,\epsilon_j]} \tau^i_{\gamma j}\,,\quad \tilde{t}^i_{\gamma j}
        :=\max_{r\in [0,\epsilon_j]} \tau^i_{\gamma j}\,.
    \end{equation*}
The ranges of values (the segments $\mathrm{ran}\,\tau^i_{\gamma
j} = [t^i_{\gamma j},\tilde{t}^i_{\gamma j}]\subset \mathbb{R}$)
for different pairs of indices $i,j$ and $i',j'$ (but for the same
$\gamma$) can intersect \textit{only} by the endpoints (so that,
either $t^i_{\gamma j}=\tilde{t}^{i'}_{\gamma j'}$ or
$\tilde{t}^i_{\gamma j}={t}^{i'}_{\gamma j'}$ holds). There is an
equality that relates functions $\tau^i_{\gamma j}$ to the
absolutely continuous spectrum of eikonals:
    \begin{equation*}
        \sigma_{\mathrm{ac}} (E_{\gamma}) = \bigcup_{j=1}^J \bigcup_{i=1}^{n_{\gamma j}}\, \mathrm{ran} \,
        \tau^i_{\gamma j}\,.
    \end{equation*}
The operators $P_{\gamma j}^i$ are {\it one-dimensional}
projectors in ${\bf l}_2^j$, mutually orthogonal for each fixed
vertex $\gamma$: $P_{\gamma j}^iP_{\gamma
j}^k=\delta_{ik}P_{\gamma j}^i$. Remarkably and importantly, in
(\ref{UEU-1 first}) they remain constant -- they do not depend on
$r\in(0,\epsilon_j)$.

The summands
\begin{equation*}
\left[{U}E_{\gamma}{U}^{-1}\right]^j :=
\sum\limits_{i=1}^{n_{\gamma j}}
        \tau_{\gamma j}^i(\cdot_j) { P_{\gamma j}^i} \in
        C\left([0,\epsilon_j];\mathfrak B({\bf
            l}^j_2)\right)
\end{equation*}
in (\ref{UEU-1 first}) are the blocks of the eikonal $E_\gamma$ in
the parametric representation.
\smallskip

\noindent$\bullet\,\,\,$ From the definition
$\mathfrak{E}_{\Sigma}:=\vee\{E_\gamma\,|\,\,\gamma\in\Sigma\}$
and (\ref{UEU-1 first}) one can get the representation
{\begin{align}
\notag & { {U}\mathfrak{E}_{\Sigma} {U}^{-1} 
{\subset}\, \bigoplus_{j=1}^J C\left([0,\epsilon_j];\mathfrak
B({\bf l}^j_2)\right);\quad {\rm pr}_j\,{U}\mathfrak{E}_{\Sigma}
{U}^{-1}=:\left[{U}\mathfrak{E}_{\Sigma}
{U}^{-1}\right]^j }=\\
\label{param form basic} & =
\vee\left\{\left[{U}E_{\gamma}{U}^{-1}\right]^j\,\big|\,\,\gamma\in\Sigma\right\}
\subset C\left([0,\epsilon_j];\mathfrak B({\bf
 l}^j_2)\right)
\end{align}
which we will refer to as the \textit{source parametric form} of
the eikonal algebra. In the first embedding, it is essential that
the algebra ${U}\mathfrak{E}_{\Sigma} {U}^{-1}$ can link blocks of
the algebra to the right. The characterization of these
connections is the main subject of \cite{Belishev_Kaplun_2022}.
\smallskip

{Let us repeat again that the above mentioned facts and
results are taken from
\cite{Belishev_Wada_2015,Belishev_Kaplun_2022}.}
\smallskip

\noindent$\bullet\,\,\,$ Let
\begin{equation}\label{proj P^j}
\mathbb{P}^j :=\{P_{\gamma j}^i\,|\,\,i=1,\dots, n_{\gamma
j};\,\,\,\gamma \in \Sigma\}.
\end{equation}
be the set of projectors associated with the block
$\left[{U}\mathfrak{E}_{\Sigma} {U}^{-1}\right]^j$. The equations
(\ref{param form basic}) are clarified as follows:
\begin{equation}\label{param form step 0}
{U}\mathfrak{E}_{\Sigma} {U}^{-1}
{\subset}\, \bigoplus_{j=1}^J
C\left([0,\epsilon_j];\mathfrak{P}^j\right);\qquad
\left[{U}\mathfrak{E}_{\Sigma} {U}^{-1}\right]^j
 \subset C([0,\epsilon_j];\mathfrak{P}^j),
 \end{equation}
where $\mathfrak{P}^j:=\vee\mathbb{P}^j\subset \mathfrak B({\bf
l}^j_2)\cong\mathbb{M}^{\,m_j}$.

\section{First canonical form}\label{sec*}

In \cite{Belishev_Kaplun_2022}, the source parametric form of the
eikonal algebra is transformed to some canonical form that
represents $\mathfrak{E}_{\Sigma}$ as a sum of {\it independent}
standard algebras of the type (\ref{standard alg}). Let us briefly
describe the transformation procedure.

\subsubsection*{Partitioning into blocks}

\noindent$\bullet\,\,\,$ On each set of projectors $\mathbb{P}^j$
we introduce the relation $\overset{\textrm{nort}}{\sim}_0$\,\,
("not orthogonal") by the rule: $P_{\gamma
j}^{i}\overset{\textrm{nort}}{\sim}_0 P_{\gamma' j}^{i'}$ if
$P_{\gamma j}^{i} P_{\gamma' j}^{i'}\neq 0$. Then we define its
transitive closure $\overset{\textrm{nort}}{\sim}$.

Let $[P]$ be the equivalence class of the projector
$P\in\mathbb{P}^j$ with respect to
$\overset{\textrm{nort}}{\sim}$. It can be
shown\,\cite{Belishev_Kaplun_2022} that the splitting
\begin{equation}\label{P irreduce}
    \mathbb{P}^j=[P]^j_1\,\cup\,\dots\cup\,[P]^j_{p_j}
\end{equation}
corresponds to the decomposition of the algebra $\mathfrak{P}^j$
into \textit{unreducible} blocks $\mathfrak{P}_p^j$:
    \begin{equation}\label{P alg to irr blocks}
        \mathfrak{P}^j = \bigoplus_{p=1}^{p_j} \mathfrak{P}_p^j\,,
    \end{equation}
    where $\mathfrak{P}_p^j:=\vee [P]^j_p\,\cong\mathbb
    M^{\,\kappa^j_p}$,\,\,\,\,$\kappa^j_1+\dots+\kappa^j_{p_j} \leqslant m_j$.
\smallskip

\noindent$\bullet\,\,\,$ In accordance with (\ref{P irreduce}),
eikonal blocks in {(\ref{param form step 0})} are decomposed
into subblocks:
    \begin{align}
        \notag &\left[UE_\gamma
        U^{-1}\right]^j=\oplus\sum\limits_{p=1}^{p_j}\left[UE_\gamma
        U^{-1}\right]^j_p,\quad \left[UE_\gamma U^{-1}\right]^j_p:=
        \sum\limits_{P^i_{\gamma j}\in[P]^j_p}\tau_{\gamma j}^i(\cdot_j)
        P_{\gamma j}^i\in \\
        \label{subblocks} & \in
        C\left([0,\epsilon_j];\mathfrak{P}_p^j\right),
    \end{align}
and the eikonal algebra satisfies
    \begin{equation*}
        {U}\mathfrak{E}_{\Sigma} {U}^{-1} \subset
        \bigoplus_{j=1}^J\left[\bigoplus\limits_{p=1}^{p_j}
         C([0,\epsilon_j],\mathfrak{P}^j_p)\right].
    \end{equation*}
Simplifying the notations, let us go to a through
 numbering of blocks, algebras and parameters:
    \begin{align*}
        & [UE_\gamma U^{-1}]^1_1,\dots,[UE_\gamma
        U^{-1}]^1_{p_1};\,\dots\,\,\dots\,;\, [UE_\gamma
        U^{-1}]^J_1,\dots,[UE_\gamma U^{-1}]^J_{p_J}\quad\to\\
        & \to\,[UE_\gamma U^{-1}]_1, \,\dots\,,[UE_\gamma U^{-1}]_L,\quad\gamma\in\Sigma\,;\\
& [P]^1_1,\dots,[P]^1_{p_1};\,\dots\,\,\dots\,;\,
        [P]^J_1,\dots,[P]^J_{p_J}\quad\to\quad[P]_1, \,\dots\,,[P]_L\,;\\
        & \mathfrak P^1_1,\dots,\mathfrak P^1_{p_1};\,\dots\,\,\dots\,;\,
        \mathfrak  P^J_1,\dots,\mathfrak P^J_{p_J}\quad\to\quad\mathfrak
        P_1, \,\dots\,,\mathfrak P_L\,;\\
        & \epsilon_1,\dots,\epsilon_{m_1};\dots\,\,\,\dots;
        \epsilon_{J-m_J},\dots,\epsilon_{J}\quad\to\quad
        \epsilon_1,\dots,\epsilon_L
    \end{align*}
and rewrite the last relation in the form
\begin{align}
\notag & {U}\mathfrak{E}_{\Sigma} {U}^{-1} \subset
\bigoplus_{l=1}^L
 C\left([0,\epsilon_l],\mathfrak{P}_l\right);\quad { {\rm pr}_lU\mathfrak{E}_{\Sigma}
U^{-1}=:}\,[U\mathfrak{E}_{\Sigma}
U^{-1}]_l=\\
\label{param form step 1} & =\vee\left\{[UE_\gamma
U^{-1}]_l\,\,\big|\,\,\gamma\in\Sigma\right\},\quad [UE_\gamma
U^{-1}]_l = \sum\limits_{P_{\gamma l}^k\in[P]_l}
        \tau_{\gamma l}^k (\cdot_l) P_{\gamma l}^k \in  C\left([0,\epsilon_l],\mathfrak{P}_l\right)
\end{align}
with \textit{irreducible} $\mathfrak {P}_l$.

\subsubsection*{Connecting the blocks}
The algebra ${U}\mathfrak{E}_{\Sigma} {U}^{-1}$ in (\ref{param
form basic}) consists of new (with respect to (\ref{param form
basic})) blocks $\left[{U}\mathfrak{E}_{\Sigma}
{U}^{-1}\right]_l:=\vee\left\{[UE_\gamma
U^{-1}]_l\,\,|\,\,\gamma\in\Sigma\right\}$.  A further
transformation of the decomposition (\ref{param form step 1}) is
possible: it consists in merging some of these blocks. Let us
briefly describe the corresponding procedure; see
\cite{Belishev_Kaplun_2022} for details.
\smallskip

\noindent$\bullet\,\,\,$ Let us choose an element
$e\in\mathfrak{E}_{\Sigma}$. Turning to the parametric form in
accordance with (\ref{param form step 1}) we have the
decomposition:
    $$
    UeU^{-1}\,=\,\oplus\sum\limits_{l=1}^L\,[UeU^{-1}]_l,\quad
    [UeU^{-1}]_l\in C([0,\epsilon_l];\mathfrak {P}_l).
    $$
The irreducibility of algebras $\mathfrak {P}_l$ implies the
irreducibility of representations of the eikonal algebra of the form
    \begin{equation*}
        \pi_l^r:\,\mathfrak{E}_{\Sigma}\to \mathfrak
        {P}_l,\quad\pi^r_l(e)\,:=\,[UeU^{-1}]_l(r),\quad 0<r<\epsilon_l
    \end{equation*}
(See \cite{Belishev_Kaplun_2022} and Proposition \ref{Prop p as
px}).  At the same time, the \textit{boundary representations}
    \begin{align}\label{boundary repr}
        \rho_l^{\pm} \mathfrak{E}_{\Sigma} \to \mathfrak{P}_l;\quad
        \rho_l^{-}(e):= \left[UeU^{-1}\right]_l(0),\quad
        \rho_l^{+}(e):=\left[UeU^{-1}\right]_l(\epsilon_l),
    \end{align}
may, in general, turn out to be reducible. In the same time, the
representations $\rho_l^{-}$ and $\rho_l^{+}$ corresponding to
{\it the same} block are obviously not equivalent. The reason for
this is that, due to the monotonicity of the functions
$\tau_{\gamma l}^j$ (see (\ref{functions tau})), the eikonal
matrices $\sum\limits_{k=1}^{n_{\gamma l}} \tau_{\gamma l}^k(0)
P_{\gamma l}^j$ and $\sum\limits_{k=1}^{n_{\gamma l}} \tau_{\gamma
l}^k(\epsilon_l) P_{\gamma l}^j$ do have {\it distinct} sets of
eigenvalues $ \{\tau_{\gamma l}^k(0)|\, \,k=1,\dots,n_{\gamma
l}\}\not=\{\tau_{\gamma l}^k(\epsilon_l)|\,\,k=1,\dots,n_{\gamma
l}\}$, which excludes equivalence.
\smallskip

\noindent$\bullet\,\,\,$ We say that the blocks
$\left[U\mathfrak{E}_{\Sigma}U^{-1}\right]_{l}$ and
$\left[U\mathfrak{E}_{\Sigma}U^{-1}\right]_{l'}$ can be connected,
if there exist representations $\rho \in
\{\rho_l^{-},\rho_l^{+}\}$ and $\rho' \in
\{\rho_{l'}^{-},\rho_{l'}^{+}\}$ which are equivalent:
$\rho\sim\rho'$.  It can be shown
\cite{Belishev_Kaplun_2022,Kaplun_dissert} that the complete set
of blocks $\{ \left[U\mathfrak{E}_{\Sigma}U^{-1}\right]_{l}| \,\,
l=1,\dots, L\}$ uniquely divides into chains of connectable ones,
whereas the order of the latter in each chain is also uniquely
determined.

Let the blocks
$\left[U\mathfrak{E}_{\Sigma}U^{-1}\right]_{l_1}$,\dots,$\left[U\mathfrak{E}_{\Sigma}U^{-1}\right]_{l_n}$
form a chain of connected blocks with the relations
$\rho_{l_1}^{+} \sim \rho_{l_2}^{-}$, $\rho_{l_2}^{+} \sim
\rho_{l_3}^{-}$,\dots, $\rho_{l_{n-1}}^{+} \sim \rho_{l_n}^{-}$
(for another order of chain connections, the consideration is
quite similar). Then there is a set of isomorphisms
$\mathbf{Y}_{l_{i}l_{i+1}}:\mathfrak{P}_{l_{i+1}}\to\mathfrak{P}_{l_i}$
which satisfy the relations
\begin{equation*}
\left[UE_{\gamma}U^{-1}\right]_{l_{i}}(\epsilon_{l_i}) =
\mathbf{Y}_{l_{i}l_{i+1}}\left(\left[UE_{\gamma}U^{-1}\right]_{l_{i+1}}(0)\right),
\quad \gamma \in\Sigma.
\end{equation*}
Denote
\begin{align*}
\mathbf{Y}_{l_{1}l_{i}}:=\mathbf{Y}_{l_{1}l_{2}}\dots\mathbf{Y}_{l_{i-1}l_{i}},\quad
r_i(r):= r - \epsilon_{l_1}-\dots - \epsilon_{l_{i-1}},\qquad
i=2,\dots, n.
\end{align*}
By the union of the chain of blocks
$\left[U\mathfrak{E}_{\Sigma}U^{-1}\right]_{l_1}$,\dots,$\left[U\mathfrak{E}_{\Sigma}U^{-1}\right]_{l_n}$
we name the algebra $\left[UE_{\gamma}U^{-1}\right]_{l_1\dots
l_n}$ defined by the equality
\begin{equation}\label{skleika def}
\left[U\mathfrak{E}_{\Sigma}U^{-1}\right]_{l_1\dots l_n} := \vee
\{E_{\gamma}^{l_1\dots l_n}|,\,\gamma \in \Sigma\}\subset
C([0,\epsilon_{l_1}+\dots+\epsilon_{l_n}];\mathfrak{P}_{l_1})
\end{equation}
with matrix-functions
\begin{equation*}
E_{\gamma}^{l_1\dots l_n}(r){:=}\left\{\begin{array}{ll}
\left[UE_{\gamma}U^{-1}\right]_{l_1}(r), & r\in[0,\epsilon_{l_1}); \\
\mathbf{Y}_{l_{1} l_{2}}\left[UE_{\gamma}U^{-1}\right]_{l_2}(r_2(r)), & r_2(r)\in[0,\epsilon_{l_2});\\
\dots & \\
\mathbf{Y}_{l_{1}
l_{n}}\left[UE_{\gamma}U^{-1}\right]_{l_n}(r_n(r)), &
r_n(r)\in[0,\epsilon_{l_n}].
\end{array}\right.
\end{equation*}
These functions are the parts of eikonals $E_\gamma$(in parametric
representation), corresponding to the new (enlarged) block formed
by the union of blocks
$\left[U\mathfrak{E}_{\Sigma}U^{-1}\right]_{l_1}$, \dots,
$\left[U\mathfrak{E}_{\Sigma}U^{-1}\right]_{l_n}$. It is also easy
to observe that the merging of blocks leads to the representation
    \begin{equation*}
        E_{\gamma}^{l_1\dots l_n}(r)=\sum\limits_{k=1}^{n_{\gamma}} \tau_{\gamma}^k(r) P_{\gamma} ^k,
    \end{equation*}
in which $n_{\gamma}: = n_{\gamma l_1}=\dots = n_{\gamma l_n}$,
$P_{\gamma} ^k : = P_{\gamma l_1} ^k $, and the functions
$\tau_{\gamma}^k$ are the extensions of linear functions
$\tau_{\gamma l_1}^k,\,k=1,\dots,n_{\gamma}$ to the larger segment
$[0,\epsilon_{l_1}+\dots,+\epsilon_{l_n}]$. Thus the possible
distinction between the union
$\left[U\mathfrak{E}_{\Sigma}U^{-1}\right]_{l_1\dots l_n}$ and the
algebra, in which it is embedded (see (\ref{skleika def})), is
again that the elements of the union may satisfy additional
conditions at the endpoints of the total segment, while the
elements of the algebra
$C([0,\epsilon_{l_1}+\dots+\epsilon_{l_n}];\mathfrak{P}_{l_1})$ do
not have them.

\subsubsection*{Canonical form}
\noindent$\bullet\,\,\,$ By performing all possible chain unions,
we present the algebra of ekonals as a sum of blocks of the form
$\left[UE_{\gamma}U^{-1}\right]_{l_1\dots l_n}$, which are no
longer connectable and are (in a relevant sense) independent
\cite{Belishev_Kaplun_2022}. Using the relations $\mathfrak{P}_l
\cong \mathbb{M}^{\kappa_l}$ we obtain that the result of such
"reformatting"\,of the source parametric representation is the
following statement which is the main subject of
\cite{Belishev_Kaplun_2022}:
    \begin{theorem}\label{Th FINAL}
There exists an isomorphism $\mathbf I$ which provides the algebra
${}{\mathfrak E}_{\Sigma}$ and its generators-eikonals representations
\begin{equation}\label{Eq FINAL}
            \mathbf I {\mathfrak E}_{\Sigma} ={\bigoplus\limits_{l=1}^\mathcal
            L} \dot C([0,\varepsilon_l];\mathbb{M}^{\,\kappa_l});\quad \mathbf{
                I} E_{\gamma} = \oplus\sum\limits_{l=1}^\mathcal
            L\left[\sum\limits_{k=1}^{n_{\gamma l }}\tau_{\gamma l}^k
            P_{\gamma l}^k\right], \,\,\, \gamma\in\Sigma\,.
\end{equation}
Here $\tau_{\gamma l}^k$ are linear functions of
$r_l\in[0,\varepsilon_l]$ such that $\big|\frac{d \tau_{\gamma
l}^k}{dr_l}\big|=1$. Their ranges $\psi_{\gamma l}^k:={\rm
ran\,}\tau_{\gamma l}^k$ are segments of length $\varepsilon_l$,
which can only share common endpoints for the same $\gamma$ and
distinct $k,l$. In this case, for all $\gamma\in\Sigma$ the
equality
        \begin{equation*}
            \sigma_{\rm ac}({E}_{\gamma})=\bigcup\limits_{l=1}^{\mathcal{L}}\bigcup\limits_{k=1}^{n_{\gamma
                    l}}\psi_{\gamma l}^k\,.
        \end{equation*}
holds.  The matrices $P_{\gamma l}^k\in \mathbb{M}^{\,\kappa_l}$
are one-dimensional projectors, mutually orthogonal for each
$\gamma$ and such that $\vee\{ P_{\gamma
l}^k\,|\,\,k=1,...\,,n_{\gamma l}
;\,\,\gamma\in\Sigma\}=\mathbb{M}^{\,\kappa_l}$.
    \end{theorem}
The isomorphic copy $\mathbf{I}\mathfrak{E}_{\Sigma}$ of the
algebra $\mathfrak{E}_{\Sigma}$ we will call its \textit{first
canonical form}. The passage to this form reveals the block
structure of the eikonal algebra.

The representation of the algebra in the form (\ref{Eq FINAL}) is
not unique, but it can be shown that any two of such
representations differ from each other only by block numbering,
their parameterization (direction of change of $r_l$) and by
replacements $P_{\gamma l}^k\to I_{l}P_{\gamma l}^k$ where $I_{l}:
\mathbb{M}^{\,\kappa_l}\to\mathbb{M}^{\,\kappa_l}$ is an
isomorphism. The functions $\tau_{\gamma l}^k$ are the same in all
representations, i.e. they are {\it invariants} of the algebra
$\mathfrak{E}_{\Sigma}$. Later on, this will allow us to use them
as coordinates on the spectrum of the eikonal algebra.
\smallskip

\noindent$\bullet\,\,\,$ Let us recall that all considerations are
performed under the assumption that the final time moment $t=T$ in
the dynamical system (\ref{Eq DS 1})--(\ref{Eq DS 4}) is fixed. As
it increases, the structure of representations (\ref{Eq FINAL})
changes. Significant changes occur at those $T$ at which the waves
propagating from the controlling vertices of $\gamma\in\Sigma$
capture new (internal or boundary) vertices: see
\cite{Belishev_Kaplun_2018, Kaplun_2021}. The evolution of eikonal
algebra over time is a separate interesting topic.

\subsubsection*{Coordinates on the spectrum}
\noindent$\bullet\,\,\,$ Due to (\ref{Eq FINAL}) and Proposition
\ref{Prop p as px}, the spectrum of the algebra ${\mathfrak
E}_{\Sigma}$ is the union of the spectra of individual standard
algebras $\dot C([0,\varepsilon_l];\mathbb{M}^{\,\kappa_l})$:
    \begin{equation*}
        \widehat{{\mathfrak E}_{\Sigma}} = {\mathcal{S}}_1\sqcup\,...\,
        {\sqcup\,{\mathcal{S}}}_\mathcal L
    \end{equation*}
(see (\ref{standard alg}), (\ref{Eq boundary hom}). Each component
({\it segment}) ${\mathcal{S}}_l$ consists of the set ({\it
interval}) $\mathrm{int}\,{\mathcal{S}}_l$ containing points which
have neighborhoods, homeomorphic to open intervals in $\mathbb{R}$
(we call them \textit{internal}), and two \textit{boundaries}
$\mathcal{K}_l^{-}$ and $\mathcal{K}_l^{+}$:
\begin{equation*}
        {\mathcal{S}}_l = \mathcal{K}_l^{-}\, \sqcup \,\mathrm{int}\,{\mathcal{S}}_l\,
\sqcup\,\mathcal{K}_l^{+}
\end{equation*}
The sets $\mathrm{int}\,{\mathcal{S}}_l$ are homeomorphic to the
corresponding intervals $(0,\varepsilon_l)$. Through $\mathrm{int}
\widehat{\mathfrak{E}_{\Sigma}}$ we denote the set of all internal
points of the spectrum. Boundaries $\mathcal{K}_l^{\pm}$ consist
of a finite number of points. We say that a boundary is a
\textit{cluster} if it contains more than one point. The points
forming the cluster are inseparable from each other in Jacobson's
topology \cite{Arveson}.

The intervals ${\rm int\,}\mathcal{S}_l$ can be metricized. As can
be easily seen from the second expression in (\ref{Eq FINAL}),
each point $\hat{\pi}\in{\rm int\,}\mathcal{S}_l$ corresponds to a
unique value of the parameter $r\in(0,\varepsilon_l)$. The
definition
\begin{equation}\label{metr int}
\delta(\hat{\pi},\hat{\pi}')\,:=\,|r-r'|,\qquad
\hat{\pi},\hat{\pi}'\in \,{\rm int\,}\mathcal{S}_l
\end{equation}
provides the natural metric on the interval. One can also
determine the distance between a boundary point and an interior
point by continuity. However, in this case the distances
(\ref{metr int}) between points in the same cluster will be zero,
since they all correspond to the same $r=0$ or $r=\varepsilon_l$.
\smallskip

\noindent$\bullet\,\,\,$ The following well-known fact motivates
the further considerations. Let $\mathfrak A$ be a
\textit{commutative} Banach algebra with a finite number of
generators $E_1,\dots,E_n$, $\widehat{\mathfrak A}$ is its
spectrum consisting of homomorphisms (characters) $\pi:\mathfrak
A\to\mathbb C$. Then the correspondence
    \begin{equation}\label{C1}
        \widehat{\mathfrak
            A}\ni\pi\,\mapsto\{\pi(E_1),\dots,\pi(E_n)\}\in\mathbb C^n
    \end{equation}
provides coordinates on the spectrum (see, e.g., \cite{Naimark}).

Let us provide an analogue of such coordinates on the spectrum of
the eikonal algebra. Technically it is more complicated, which is
to be expected since $\mathfrak E_\Sigma$ is noncommutative.
\smallskip

\noindent$\bullet\,\,\,$ Let
$\hat{\pi}\in\widehat{\mathfrak{E}}_\Sigma$ be an arbitrary point
of the spectrum of the eikonal algebra and let $\pi\in\hat{\pi}$
be some of its representatives (irreducible representation in a
Hilbert space $H_{\pi}$). For each vertex $\gamma\in\Sigma$ we
define the operator $e_{\gamma}(\pi)$ by
    \begin{equation*}
        e_{\gamma}(\pi):=\pi (E_{\gamma}) \in
        \mathfrak{B}(H_{\pi})\,;
    \end{equation*}
let $\sigma^{+}(e_{\gamma}(\pi))$ be the set of its positive
eigenvalues. In this case there is equality
    \begin{equation}\label{sigma + = sigma +}
        \sigma^{+}(e_{\gamma}(\pi))=\sigma^{+}(e_{\gamma}(\pi'))
    \end{equation}
for any representations $\pi,\pi'\in\hat{\pi}$ from the same
equivalence class.

Let $\hat{\pi}\in \mathrm{int} \widehat{\mathfrak{E}_{\Sigma}}$ be
an interior point of the spectrum. For it, we define {\it
$\gamma$-coordinates} by the equality
    \begin{equation*}
        \sigma_{\gamma}(\hat{\pi}) := \sigma^{+}(e_{\gamma}(\pi)),\qquad
        \pi {\in}\hat{\pi}\,;
    \end{equation*}
the correctness of the definition comes from (\ref{sigma + = sigma
+}). From the second expression in (\ref{Eq FINAL}) and the
properties of its included functions $\tau^k_{\gamma l}$, it is
easy to see that the selected point $\hat{\pi}$ corresponds to a
unique number $l$ and parameter $r\in(0,\varepsilon_l)$ such that
the following equality holds
    \begin{equation*}
        \sigma_{\gamma}(\hat{\pi}) = \{\tau_{\gamma l}^1(r),\dots, \tau_{\gamma l}^{n_{\gamma l}}(r)\}
    \end{equation*}
For the boundary points $\hat{\pi}\in \mathcal{K}^{\pm}_l$ we take
by continuity:
    \begin{equation}\label{coord clasters}
\sigma_{\gamma} (\hat{\pi}) :=\left\{\lim_{r\to c}\tau_{\gamma
l}^1(r),\,\dots,\, \lim_{r\to c}\tau_{\gamma l}^{n_{\gamma
l}}(r)\right\}, \qquad c=0,\varepsilon_l\,.
    \end{equation}
Note that if the boundary set is a cluster, then all its points
will be assigned {\it the same} numerical set (\ref{coord
clasters}).

The correspondence
\begin{equation}\label{coord for eik alg}
\widehat{\mathfrak{E}}_\Sigma\ni\hat{\pi}\,\mapsto\left\{\sigma_{\gamma}(\hat{\pi})\,|\,\,\gamma\in\Sigma\right\}
\end{equation}
is proposed as a generalization (\ref{C1}) for the case of
\textit{noncommutative} eikonal algebras.

It should be mentioned that the sets in the right-hand side of
(\ref{coord for eik alg}) are not coordinates in the rigorous
sense: as noted, they distinguish internal points of the spectrum
but do not distinguish points belonging to the same cluster.
Nevertheless, they are useful because they provide true
coordinates on the set $\mathrm{int}
\widehat{\mathfrak{E}_{\Sigma}}$, i.e., on the main part of the
spectrum of algebra ${\mathfrak{E}_{\Sigma}}$.

\subsubsection*{Frame ${\mathfrak F}^{\,\rm a}_\Sigma$}

Here we introduce some equivalence relation for points of
$\widehat{\mathfrak{E}}_\Sigma$. Factorization (gluing) of the
spectrum by this relation will turn it into a graph.
\smallskip

\noindent$\bullet\,\,\,$ Define on $\widehat{\mathfrak{E}}_\Sigma$
the relation $\sim_0$ by the following rule:
$\hat{\pi}\sim_0\hat{\pi}'$ if there exists a vertex
$\gamma\in\Sigma$ such that
$\sigma_{\gamma}(\hat{\pi})\cap\sigma_{\gamma}(\hat{\pi}')\neq
\emptyset$. Let $\sim$ be the transitive closure of $\sim_0$ and
$[\hat{\pi}]$ be the equivalence class of the spectrum point
$\hat{\pi}$.
\begin{propos}\label{Lemma about factor}
Let $\hat{\pi}\in\widehat{\mathfrak{E}}_\Sigma$ be a point of the
spectrum, $[\hat{\pi}]$ is its equivalence class. Then:

\noindent{{{1.}}} if
$\hat{\pi}\in\mathrm{int}\,\widehat{\mathfrak{E}}_\Sigma$ is an
interior point, then $[\hat{\pi}]=\{\hat{\pi}\}$, i.e., its
equivalence class is limited to the point itself;

\noindent{{{2.}}} if $\hat{\pi}\in\mathcal{K}^\pm_l$ is a point of
a boundary set (possibly a cluster), then there is an embedding
$\mathcal{K}^\pm_l\subset[\hat{\pi}]\subset
\widehat{\mathfrak{E}}_\Sigma\setminus
\mathrm{int}\,\widehat{\mathfrak{E}}_\Sigma$.
\end{propos}
Part 1 easily follows from the properties of the functions
$\tau_{\gamma l}^k$ (see Theorem \ref{Th FINAL}), namely, the
disjunction of their ranges $\psi_{\gamma l}^k$. The embedding in
Part 2 follows directly from the definition of equivalence, and
the difference $[\hat{\pi}]\setminus \mathcal{K}^\pm_l$ can
consist of the points of another boundary sets
$\mathcal{K}^\pm_{l'}$ that got into class $[\hat{\pi}]$ during
factorization.
\smallskip

\noindent$\bullet\,\,\,$ We say the factor-space ${\mathfrak
F}^{\,\rm a}_\Sigma:=\widehat{\mathfrak{E}}_\Sigma/_{\sim}$ to be
a \textit{algebraic frame} of the domain $\Omega^T[\Sigma]$. By
${\rm proj}:\, \widehat{\mathfrak{E}}_\Sigma\to {\mathfrak
F}^{\,\rm a}_\Sigma$ we denote the canonical projection. The
spectrum is equipped with the Jacobson topology and hence there is
a canonical factor-topology on ${\mathfrak F}^{\,\rm a}_\Sigma$.

Proposition \ref{Lemma about factor} implies that the passage from
the spectrum $\widehat{\mathfrak{E}}_\Sigma$ to the frame
${\mathfrak F}^{\,\rm a}_\Sigma$ is reduced to connecting some
segments $\mathcal{S}_l$ by identifying points of their boundary
sets $\mathcal{K}^\pm_l$. The clusters, which make the spectrum
non-Hausdorff space, are glued into points during factorization.
As a consequence, the space ${\mathfrak F}^{\,\rm a}_\Sigma$ turns
out to be \textit{Hausdorff}, and each of its components is
homeomorphic to some \textit{graph} whose edges are ${\rm
proj\,}({\rm int\,}\mathcal{S}_l)$ and its vertices are the
points, formed by gluing some of the boundaries together:
\begin{align*}
\notag &{\mathfrak F}^{\,\rm a}_\Sigma=\mathscr E^{\rm
a}\sqcup\mathscr W^{\rm a};\quad \mathscr E^{\rm a}=\{{\rm
proj\,}({\rm int\,}\mathcal{S}_1),\dots,{\rm proj\,}({\rm
int\,}\mathcal{S}_\mathcal L)\},\,\,\, \mathscr W^{\rm a}=\{w_1,\dots,w_p\},\\
& w_k={\rm proj\,}\mathcal K^{\alpha_1}_{l_1}=\,\dots\,={\rm
proj\,}\mathcal K^{\alpha_k}_{l_k},\quad \alpha_k\in\{-,+\}.
\end{align*}
Note that the passage from spectrum to frame may result in
appearance of the valency 2 vertices.
\smallskip

\noindent$\bullet\,\,\,$ To simplify the writing, we will denote
frame points by
$\scalemath{1.35}{\mbox{\boldmath$\pi$}}:=[\hat{\pi}]$. Let's
introduce $\gamma-$coordinates:
\begin{align*}
\label{bold sigma def} &
\scalemath{1.35}{\mbox{\boldmath$\sigma$}}_{\gamma}(\scalemath{1.35}{\mbox{\boldmath$\pi$}}):
=\bigcup_{\hat{\pi}\in\scalemath{1.35}{\mbox{\boldmath$\pi$}}}
\sigma_{\gamma}(\hat{\pi}),\qquad {
\scalemath{1.35}{\mbox{\boldmath$\pi$}}\in{\mathfrak F}^{\,\rm
a}_\Sigma}
\end{align*}
(the union of numerical sets on a common numerical axis) and
define the coordinates on the whole frame by the rule
\begin{equation}\label{sigma gamma for frame spec}
{\mathfrak F}^{\,\rm a}_\Sigma\ni
\scalemath{1.35}{\mbox{\boldmath$\pi$}}\,\mapsto\,\left\{
\scalemath{1.35}{\mbox{\boldmath$\sigma$}}_{\gamma}(\scalemath{1.35}{\mbox{\boldmath$\pi$}})\,|\,\,\gamma\in\Sigma\right\}.
\end{equation}}
From the Proposition \ref{Lemma about factor} one can easily
conclude that the sets in the right part (\ref{sigma gamma for
frame spec}) distinguish {\it all} frame points, i.e. they are the
proper coordinates on $\mathfrak{F}_{\Sigma}$.
\smallskip

\noindent$\bullet\,\,\,$ As follows from the Proposition
\ref{Lemma about factor}, the projection $\rm proj$ acts
injectively on the interior points of the spectrum. This allows us
to metricize the edges of the frame by the rule
\begin{equation*}
\Delta(\scalemath{1.35}{\mbox{\boldmath$\pi$}},\scalemath{1.35}{\mbox{\boldmath$\pi$}}')\,:=\,\delta\left({\rm
proj}^{-1}(\scalemath{1.35}{\mbox{\boldmath$\pi$}}),{\rm
proj}^{-1}(\scalemath{1.35}{\mbox{\boldmath$\pi$}}')\right)
\end{equation*}
(see (\ref{metr int})), and then, by analogy with the metric
(\ref{metric tau}) on the stars, extend the $\Delta$ metric to the
frame vertices. As a result, the whole frame ${\mathfrak F}^{\,\rm
a}_\Sigma$ turns out to be a (possibly non-connected) compact {\it
metric graph}.
\smallskip

\noindent$\bullet\,\,\,$ The result of the previous considerations
is a remarkable fact: the part of the graph $\Omega$ filled with
waves, by the scheme
$$
\Omega^T[\Sigma]\to\mathfrak{E}_{\Sigma}\to\widehat{\mathfrak{E}_{\Sigma}}\to{\mathfrak F}^{\,\rm a}_\Sigma\,,
$$
is canonically mapped to some metric canonically coordinated graph
that is the frame ${\mathfrak F}^{\,\rm a}_\Sigma$ extracted from
the algebra $\mathfrak{E}_{\Sigma}$. As we can easily see from
Theorem \ref{Th FINAL} and the remarks below it, this construction
is an {\it invariant} of the eikonal algebra: the frames
corresponding to different versions of the representation (\ref{Eq
FINAL}) are isometric.

\subsubsection*{Functional model}
\noindent$\bullet\,\,\,$  Let the representation (\ref{Eq FINAL})
be fixed.  Recall that then each point of the spectrum
$\hat\pi\in\widehat{\mathfrak E_\Sigma}$ is associated with a
certain value of the parameter $r_{\hat\pi}$ taking values in the
segments $[0,\varepsilon_l]$. The correspondence ${\hat\pi}\to
r_{\hat\pi}$ is injective on the set of interior points ${\rm
int\,}{\mathfrak E_\Sigma}$.

Let $[\mathbf{I} e]_l(\cdot) \in \dot C
([0,\varepsilon_l];\mathbb{M}^{\kappa_l})$ be the $l-$th block of
the element $\mathbf{I} e$ in the form (\ref{Eq FINAL}). For each
point $\hat{\pi} \in \mathrm{int\,} \mathcal{S}_l$ we choose a
representative $\pi \in \hat{\pi}$ such that equality
$\pi(e)=[\mathbf{I} e]_l(r_{\hat{\pi}})$ holds. For the points of
the boundary sets $\mathcal{K}_l^{\pm}$ we fix their numbering
$\mathcal{K}_l^{\pm}= \{\hat{\pi}_1^{\pm},\dots,
\hat{\pi}^{\pm}_{m_l^{\pm}}\}$ and representatives
$\pi_k^{\pm}\in\hat{\pi}_k^{\pm}$ such that the equality
$\oplus\sum\limits_{k=1}^{m_l^{\pm}} \pi_k^{\pm} (e) = [\mathbf{I}
e]_l(c^{\pm}) $,\quad $c^{+}=\epsilon_l$, $c^{-}=0$ holds. Then,
matrix-functions $\mathbf{I}e $ are transferred to the spectrum by
the rule
\begin{equation*}
e(\hat\pi):= \pi (e),\quad \pi \in \hat{\pi} \in
\widehat{\mathfrak{E}_{\Sigma}},
\end{equation*}
where $\pi $ is a representative of $\hat{\pi}$ defined above.

These functions are then transferred from the spectrum to the
frame $\mathfrak{F}^{\,\rm a}_{\Sigma}$ by the rule
\begin{equation}\label{func mod 1}
e({\mbox{\boldmath$\pi$}}):=
 \begin{cases}
e({\rm proj}^{-1}(\scalemath{1.35}{\mbox{\boldmath$\pi$}})), &
\scalemath{1.35}{\mbox{\boldmath$\pi$}}\in\mathscr E^{\rm a}; \cr
 \oplus\sum\limits_{{\hat\pi}_k\in{\rm
proj}^{-1}(\scalemath{1.35}{\mbox{\boldmath$\pi$}})} e({\hat\pi}_k),&
\scalemath{1.35}{\mbox{\boldmath$\pi$}}\in\mathscr W^{\rm a}.
  \end{cases}
\end{equation}
In the second line in (\ref{func mod 1}) there is a matrix
composed of blocks arranged in some order. The order is not
crucial, but it is assumed that for each
$\scalemath{1.35}{\mbox{\boldmath$\pi$}}\in\mathscr W^{\rm a}$ the
order is fixed.

As a result, the eikonal algebra is realized as an algebra of of
matrix-valued functions on an algebraic frame. These functions are
continuous on its edges and, generally speaking, discontinuous at
vertices.
\smallskip

\noindent$\bullet\,\,\,$ The above construction can be interpreted
as a C*-algebra bundle over the base $\mathfrak{F}^{\,\rm
a}_{\Sigma}$, and the functions (\ref{func mod 1}) as its sections
\cite{Vas,Nilsen_1996}. It can be shown that these sections have
an additional property {\it semicontinuity}: see
\cite{Nilsen_1996}.

\section{Second canonical form}\label{sec3}

Converting the eikonal algebra to the second form starts from the
same source parametric form (\ref{param form basic}), (\ref{param
form step 0}).

\subsubsection*{Partitioning into blocks}
\noindent$\bullet\,\,\,$ Recall that the projectors $P_{\gamma
j}^i$ in the source parametric form are one-dimensional operators
which act in ${\bf l}^j_2$. In the indicator basis they take the
form
\begin{equation*}
 P_{\gamma j}^i=(\,\cdot\,,\beta^i_{\gamma j})_{{\bf
 l}_2^j}\,\beta^i_{\gamma j},\quad \text{where}\quad\beta^i_{\gamma
 j}=\sum\limits_{l=1}^{m_j} \beta^{il}_{\gamma j}\,\chi_k\in {\bf
 l}^j_2,\quad (\beta^i_{\gamma j},\beta^{k}_{\gamma j})_{{\bf
 l}_2^j}=\delta_{ik},
\end{equation*}
and get the matrices $\check p^{\,i}_{\gamma
j}=\left\{\beta^{il}_{\gamma j}\beta^{il'}_{\gamma
j}\right\}_{l,l'=1}^{m_j}\in\mathbb M^{\,m_j}$. Just like the
projectors $P_{\gamma j}^i$\, the vectors $\beta_{\gamma j}^i\in
{\bf l}_2^j$ that represent them, do not depend on $r_j$. If we
consider the elements of the space ${\bf l}^j_2$ as {numerical
functions} on the set $\Lambda[x(r)]\subset\Phi^j$ (see (\ref{unit
U^j})), then for each vector its {\it support} ${\rm
supp\,}\beta_{\gamma j}^i\subset\Lambda[x(r)]$ is defined. The
definition is correct because both the projectors $P_{\gamma
j}^i$, and the vectors $\beta_{\gamma j}^i$ associated with them,
do not depend on $r\in(0,\epsilon_j)$. Let us note the equivalence
$$
 {\rm supp\,}\beta_{\gamma j}^i\,\cap\, {\rm supp\,}\beta_{\gamma'
 j}^{i'}\not=\emptyset\quad\Leftrightarrow
 \quad\sum\limits_{l=1}^{m_j} |\beta^{il}_{\gamma
 j}|\,|\beta^{i'l}_{\gamma' j}|\not=0,\qquad
 \gamma,\gamma'\in\Sigma\,
$$
and put by definition: \,\,\,${\rm supp\,}P_{\gamma j}^i:={\rm
supp\,}\beta_{\gamma j}^i$.
\smallskip

\noindent$\bullet\,\,\,$ Now, on each set of projectors
$\mathbb{P}^j$ (see (\ref{proj P^j})), we define the relation
$\overset{\textrm{supp}}{\sim}_0$\,\, by the following rule:
$P_{\gamma j}^{i}\overset{\textrm{supp}}{\sim}_0 P_{\gamma'
j}^{i'}$ if ${\rm supp\,}P_{\gamma j}^{i}\,\cap\, {\rm supp\,}
P_{\gamma' j}^{i'}\not=\emptyset$. By
$\overset{\textrm{supp}}{\sim}$ we denote its transitive closure.

Let $\langle P\rangle$ be the equivalence class of the projector
$P\in\mathbb{P}^j$ with respect to
$\overset{\textrm{supp}}{\sim}$. From the definition of the
relation $\overset{\textrm{supp}}{\sim}$ it easily follows that
the partition
\begin{equation}\label{P irreduce+} \mathbb{P}^j=\langle
P\rangle^j_1\,\cup\,\dots\cup\,\langle P\rangle^j_{q_j}
\end{equation}
leads to the decomposition of the algebra $\mathfrak{P}^j=\vee
\mathbb P^j$ into orthogonal blocks:
\begin{equation}\label{P alg to irr blocks+}
\mathfrak{P}^j = \bigoplus_{q=1}^{q_j} \mathfrak{Q}_q^j\,,\qquad
\mathfrak{Q}_q^j:=\vee \langle P\rangle^j_q\,.
\end{equation}
\noindent$\bullet\,\,\,$ It is useful to compare the
representations (\ref{P irreduce+}) and (\ref{P alg to irr
blocks+}) with (\ref{P irreduce}) and (\ref{P alg to irr blocks}):
if the algebras $\mathfrak{P}_p^j\cong\mathbb M^{\,\kappa^j_p}$
are irreducible, then the irreducibility of $\mathfrak{Q}_q^j$ is
generally speaking not guaranteed. At the same time, one can show
\cite{Kaplun_dissert} that the representation (\ref{P irreduce+})
corresponds to the partition
\begin{equation*}
\Lambda[x(r)]\,=\,\Lambda^j_1[x(r)]\,\cup\,\dots\,\cup\,\Lambda^j_{q_j}[x(r)]\subset\Phi^j
\end{equation*}
on {\it minimal} (by number of points) determination sets
$\Lambda^j_q[x(r)]=\{x_{1}^{j,q}(r),\dots,x_{s_q}^{j,q}(r)\}$,
i.e., on the smallest sets having the property (\ref{det set}).
Correspondingly, there is a decomposition of families into
subfamilies:
\begin{equation}\label{new Phi}
\Phi^j=\Phi^j_1\cup\dots\cup\Phi^j_{q_j},\,\,\,
\Phi^j_{q}=\bigcup\limits_{0<r<\epsilon_j}\Lambda^j_q[x(r)]=\bigcup\limits_{s=1}^{s_q}\omega^{j}_{k(s)},\,\,\,
\omega^{j}_{k(s)}=\bigcup\limits_{0<r<\epsilon_j}x_{k(s)}^{j,q}(r)\,.
\end{equation}
With the latter in mind, we can say that the partition (\ref{P
irreduce+}), unlike (\ref{P irreduce}), has a geometric meaning:
it corresponds to a graph partition
\begin{equation*}
\overline{\Omega^T[\Sigma]}=\left[\bigcup\limits_{\substack{j=1,...,J;\\q=1,...,q_j}}\Phi^j_q\right]\cup\Theta\,.
\end{equation*}
In this case, by construction of the representation (\ref{new
Phi}), the following is fulfilled
\begin{equation}\label{L_2 decomp}
L_2(\overline{\Omega^T[\Sigma]})=\oplus\sum\limits_{j=1}^J\sum\limits_{q=1}^{q_j}L_2(\Phi^j_q);\quad
E_\gamma L_2(\Phi^j_q)\subset
L_2(\Phi^j_q),\,\,\,\gamma\in\Sigma\,.
\end{equation}

\noindent$\bullet\,\,\,$ According to (\ref{L_2 decomp}), the
blocks of eikonals in (\ref{param form step 0}) are decomposed
into subblocks:
\begin{align}
\notag &\left[UE_\gamma
U^{-1}\right]^j=\oplus\sum\limits_{q=1}^{q_j}\left\langle
UE_\gamma U^{-1}\right\rangle^j_q,\quad \left\langle UE_\gamma
U^{-1}\right\rangle^j_q = \sum\limits_{P^i_{\gamma j}\in\langle
P\rangle^j_q}\tau_{\gamma j}^i(\cdot_j)
P_{\gamma j}^i\in\\
\label{subblocks} & \in
C\left([0,\epsilon_j];\mathfrak{Q}_q^j\right),
\end{align}
and for the eikonal algebra we have the following relation
\begin{equation*}
{U}\mathfrak{E}_{\Sigma} {U}^{-1} \subset
\bigoplus_{j=1}^J\left[\bigoplus\limits_{q=1}^{q_j}
C\left([0,\epsilon_j];\mathfrak{Q}^j_q\right)\right].
 \end{equation*}
Simplifying the notations, we will proceed to a through numbering:
\begin{align*}
        & \Phi^1_1,\dots, \Phi^1_{q_1};\,\dots\,\,\dots\,;\,
        \Phi^J_1,\dots,\Phi^J_{q_J}\quad\to\quad
        \Phi_1, \,\dots\,, \Phi_M\,;\\
        & \Lambda^1_1,\dots, \Lambda^1_{q_1};\,\dots\,\,\dots\,;\,
            \Lambda^J_1,\dots,\Lambda^J_{q_J}\quad\to\quad
            \Lambda_1, \,\dots\,, \Lambda_M\,;\\
& \langle P\rangle^1_1,\dots, \langle
P\rangle^1_{q_1};\,\dots\,\,\dots\,;\,
            \langle P\rangle^J_1,\dots,\langle P\rangle^J_{q_J}\quad\to\quad
            \langle P\rangle_1, \,\dots\,, \langle P\rangle_M\,;\\
        & \langle UE_\gamma U^{-1}\rangle^1_1,\dots,\langle UE_\gamma
        U^{-1}\rangle^1_{q_1};\,\dots\,\,\dots\,;\, \langle UE_\gamma U^{-1}\rangle^J_1,\dots,\langle UE_\gamma U^{-1}\rangle^J_{q_J}\quad\to\\
        & \to\,\langle UE_\gamma U^{-1}\rangle_1, \,\dots\,,\langle UE_\gamma U^{-1}\rangle_M,\quad\gamma\in\Sigma\,;\\
        & \mathfrak Q^1_1,\dots,\mathfrak Q^1_{q_1};\,\dots\,\,\dots\,;\,
        \mathfrak  Q^J_1,\dots,\mathfrak Q^J_{q_J}\quad\to\quad\mathfrak
        Q_1, \,\dots\,,\mathfrak Q_M\,;\\
        & \epsilon_1,\dots,\epsilon_{m_1};\dots\,\,\,\dots;
        \epsilon_{J-m_J},\dots,\epsilon_{J}\quad\to\quad
        \epsilon_1,\dots,\epsilon_M.
    \end{align*}
In the new notation we have:
\begin{align}
\notag
&\overline{\Omega^T[\Sigma]}=\left[\bigcup\limits_{l=1}^M\Phi_l\right]\cup\Theta;\quad
\Phi_l=\bigcup\limits_{0<r<\epsilon_l}\Lambda_l[x(r)]=\bigcup\limits_{s=1}^{m_l}\omega_{ls},\quad
\Lambda_l[x(r)]=\\
\label{new Phi+} & =\{x^l_1(r),\dots,x^l_{m_l}(r)\},\quad \omega_{ls}=\bigcup\limits_{0<r<\epsilon_l}x^l_s(r)\,;\\
\notag & {U}\mathfrak{E}_{\Sigma} {U}^{-1} 
\subset \bigoplus_{l=1}^M
C\left([0,\epsilon_l];\mathfrak{Q}_l\right);\quad {\rm
pr}_l\,{U}\mathfrak{E}_{\Sigma} {U}^{-1}  =:\langle
U\mathfrak{E}_{\Sigma}
U^{-1}\rangle_l=\\
\label{param form step 1+}& =\vee\left\{\langle UE_\gamma
U^{-1}\rangle_l\,\,\big|\,\,\gamma\in\Sigma\right\}, \,\,\,\langle
UE_\gamma U^{-1}\rangle_l = \sum\limits_{P_{\gamma l}^k \in\langle
P\rangle_l} \tau_{\gamma l}^k (\cdot_l) P_{\gamma l}^k \in
C\left([0,\epsilon_l],\mathfrak{Q}_l\right).
    \end{align}

\subsubsection*{Connecting the families}
Connecting families is a procedure similar to the connecting
blocks in the passage to the first form. The important difference
is that now all the steps remain relation to the geometry and, as
a result, the second form will correspond to a new partition of
the graph $\overline{\Omega^T[\Sigma]}$ into families $\Phi$. The
new families, generally speaking, consist of larger cells and are
distinguished by the fact that they are formed by {\it minimal}
$\Lambda_l[x(r)]$ under variations of the parameter $r$: see
(\ref{new Phi+}).
\smallskip

\noindent$\bullet\,\,\,$ As $r$ varies, the sets
$\Lambda_l[x(r)]=\{x^l_1(r),\dots,x^l_{m_l}(r)\}$ continuously
vary their position on the graph. When the parameter tends to the
boundary values $0$ or $\epsilon_l$, they go to the limiting sets
$\Lambda_l[x(0)], \,\Lambda_l[x(\epsilon_l)]\,\subset\Theta'$
consisting of boundaries $x^l_s(0),\,x^l_{s}(\epsilon_l)$ of the
cells $\omega_{l s}\subset\Phi_l$.

For the elements $UeU^{-1}\in
\langle{U}\mathfrak{E}_{\Sigma}{U}^{-1}\rangle_l\subset
C\left([0,\epsilon_l];\mathfrak{Q}_l\right)$ we define the
\textit{boundary representations} $\rho_l^{\pm}
\mathfrak{E}_{\Sigma} \to \mathfrak{Q}_l$:
\begin{equation*}
\rho_l^{-}(e):=\left\langle UeU^{-1}\right\rangle_l(0),\quad
\rho_l^{+}(e):=\left\langle
UeU^{-1}\right\rangle_l(\epsilon_l);\qquad l=1,\dots, M.
\end{equation*}
Just like representations (\ref{boundary repr}), they can
generally be reducible, and representations $\rho_l^{-}$ and
$\rho_l^{+}$ that correspond to the same block, are not
equivalent.
\smallskip

\noindent$\bullet\,\,\,$ Let $\Phi_l$ and $\Phi_{l'}$ be two
families with equal number of cells: $m_l=m_{l'}=:m_{ll'}$. We say
that these families are connectable if
\smallskip

\noindent$\bf 1.$ under some parameterization of them,
$\Lambda_l[x(\epsilon_l)]=\Lambda_{l'}[x(0)]$ is fulfilled, where
$\#\Lambda_l[x(\epsilon_l)] =\#\Lambda_{l'}[x(0)] = m_{l l'}$;
\smallskip

\noindent$\bf 2.$ under a suitable choice of cell numbering, the
bijection
$\Lambda_l[x(\epsilon_l)]\leftrightarrow\Lambda_{l'}[x(0)]$, given
by the condition $x^l_s(\epsilon_l) = x^{l'}_{s}(0)$, $s=1,\dots,
m_{ll'}$ , {is well-defined}. It determines the unitary operator
$V_{ll'}:{\bf l}_2^l\to{\bf
l}_2^{l'},\,\,V_{ll'}\chi^l_s=\chi^{l'}_s,\,\,\,s=1,\dots,m_{ll'}$\,
which links the boundary representations:
$V_{ll'}\rho^+_l=\rho^-_{l'}V_{ll'}$ (and thus establishes their
equivalence).

In this case, the cells $\omega_{ls}\subset\Phi_l$ and
$\omega_{l's}\subset\Phi_{l'}$ are coupled in pairs on the graph
through common boundaries $x^l_s(\epsilon_l)=x^{l'}_s(0)$. Such a
connection of cells will be denoted by $\omega_{ls}
\leftrightarrow \omega_{l's}$. It is not difficult to see that the
complete set of families $\{\Phi_l\,|\,\, l=1,\dots, M\}$ breaks
down into chains of connectable ones.

Let $\Phi_{l_1}$,\dots,$\Phi_{l_n}$ be a chain of connectable
families and the relations $\rho_{l_1}^{+} \sim \rho_{l_2}^{-}$,
$\rho_{l_2}^{+} \sim \rho_{l_3}^{-}$,\dots, $\rho_{l_{n-1}}^{+}
\sim \rho_{l_n}^{-}$ hold (the considerations are quite similar
for the other order of connections in the chain). Thus a chain of
cells is established:
    \begin{equation*}
\omega^{l_1}_{s} \leftrightarrow \omega^{l_2}_{s}
\leftrightarrow\dots
\leftrightarrow\omega^{l_{n-1}}_{s}\leftrightarrow\omega^{l_n}_{s},\quad
s=1,\dots,m_{l_1\dots l_n}
    \end{equation*}
where $m_{l_1\dots l_n}:= m_{l_1}=\dots = m_{l_n}$. Let us define
the cells $\omega^{l_1\dots l_n}_s$ by the equalities:
    \begin{equation*}
\omega^{l_1\dots l_n}_s:=\omega^{l_1}_{s} \cup
\overline{\omega^{l_2}_{s}}\cup\dots
\cup\overline{\omega^{l_{n-1}}_s}\cup\omega^{l_n}_{s},\quad
s=1,\dots,m_{l_1\dots l_n}.
    \end{equation*}
The set $\Phi_{l_1\dots l_n}$ defined by
    \begin{equation*}
        \Phi_{l_1\dots l_n}:=\bigcup_{i=1}^{m_{l_1\dots l_n}} \omega^{l_1\dots l_n}_i;
    \end{equation*}
is said to be the union of the families
$\Phi_{l_1}$,\dots,$\Phi_{l_n}$.

One can easily see that $\Phi_{l_1\dots l_n}$ is also a family,
and invariance of spaces $L_2(\Phi_l)$ for eikonals in (\ref{L_2
decomp}) implies invariance $E_\gamma L_2(\Phi_{l_1\dots
l_n})\subset L_2(\Phi_{l_1\dots l_n}),\,\,\,\gamma\in\Sigma$.
{Searching} the connecting procedure, it is easy to verify that the
values of the functions $\tau^i_{\gamma j}$, which correspond to
the blocks $\langle UE_\gamma U^{-1}\rangle_{l_1}$, \dots,
$\langle UE_\gamma U^{-1}\rangle_{l_n}$ in the eikonal
representation (\ref{subblocks}), are properly connected on the
cell boundaries $\omega_{l_1 s}$,\dots,$\omega_{l_n s}$, so that
they form linear functions $\tau^s_{\gamma l_1\dots l_n}$ on the
new cells $\omega_{l_1\dots l_n,\,s}$. In the mean time,
equivalence of boundary representations implies that the
projectors $P^i_{\gamma j}$ corresponding to blocks $\langle
UE_\gamma U^{-1}\rangle_{l_1}$, \dots, $\langle UE_\gamma
U^{-1}\rangle_{l_n}$, are sequentially intertwined by operators
$V_{l_i l_{i+1}}$. Hence, they have the same matrices $\check
p^s_{\gamma l_1\dots l_n}$ in the corresponding indicator bases.

As a result, in all previous representations, the chain of blocks
$\langle UE_\gamma U^{-1}\rangle_{l_1}$, \dots, $\langle UE_\gamma
U^{-1}\rangle_{l_n}$ is replaced by one block $\langle
U\mathfrak{E}_{\Sigma} U^{-1}\rangle_{l_1\dots l_n}$ -- their
union. Repeating this procedure for all chains of connectable
blocks, we arrive at a partition of the wave-filled domain
$\Omega^T[\Sigma]$ into families that {\it no longer admit
connections}:
\begin{align}
\notag
&\overline{\Omega^T[\Sigma]}=\left[\bigcup\limits_{l=1}^\mathcal
M\Phi'_l\right]\cup\Theta';\quad
\Phi'_l=\bigcup\limits_{0<r<\epsilon'_l}\Lambda'_l[x'(r)]=\bigcup\limits_{s=1}^{m'_l}\omega'_{ls},\quad
\Lambda'_l[x'(r)]=\\
\label{new Phi++} & =\{{x'}^l_1(r),\dots,{x'}^l_{m'_l}(r)\},\quad
\omega'_{ls}=\bigcup\limits_{0<r<\epsilon'_l}{x'}^l_s(r)\,.
\end{align}
Here $\Lambda'_l[x(r)]$ are the minimal determination sets, which
gives reason to call the partition (\ref{new Phi++}) {\it
optimal}.

As one can easily see, the optimal partition reduces eikonals:
$E_\gamma L_2(\Phi'_l )\subset L_2(\Phi'_l )$ holds for all
$\gamma\in\Sigma$.

\subsubsection*{Geometric form}
The previous considerations are summarized in the form of the
following analogue of Theorem \ref{Th FINAL}.
\begin{theorem}\label{Th FINAL2}
The optimal partition corresponds to the representation of the eikonal algebra in the form
\begin{align}
\notag & {U}\mathfrak{E}_{\Sigma}{U}^{-1}\subset
\bigoplus_{l=1}^\mathcal
MC\left([0,\epsilon'_l];\mathfrak{Q}'_l\right)
;\quad {\rm pr}_l\,U\mathfrak{E}_{\Sigma} U^{-1}=
\left\langle{U}\mathfrak{E}_{\Sigma}{U}^{-1}\right\rangle'_l=
\\
\notag & = \vee\left\{\left\langle UE_\gamma
U^{-1}\right\rangle'_l\,\big|\,\,\gamma\in\Sigma\right\},\quad
\left\langle UE_\gamma U^{-1}\right\rangle'_l =
\sum\limits_{s=1}^{n'_{\gamma l}}{\tau'}_{\gamma l}^{\,s}(\cdot_l)
{P'}_{\gamma l}^{\,s}\,,\\
\label{geom form} & UE_\gamma
U^{-1}=\oplus\sum_{l=1}^{\mathcal{M}}\left\langle UE_\gamma
U^{-1}\right\rangle'_l
\end{align}
with the blocks $\left\langle U\mathfrak{E}_{\Sigma}
U^{-1}\right\rangle'_l\subset
C\left([0,\epsilon'_l];\mathfrak{Q}'_l\right)$. The functions
${\tau'}_{\gamma l}^{\,s}$ are linear functions of
$r_l\in[0,\epsilon'_l]$ such that $\big|\frac{d {\tau'}_{\gamma
l}^{\,s}}{dr_l}\big|=1$. Their ranges $\xi_{\gamma l}^s:={\rm
ran\,}{\tau'}_{\gamma l}^{\,s}$ are segments of the length
$\epsilon'_l$, which may have (for the same $\gamma$ and different
$s,l$) only common endpoints. In this case, for all
$\gamma\in\Sigma$ the following equality holds
\begin{equation*}
\sigma_{\rm
ac}({E}_{\gamma})=\bigcup\limits_{l=1}^{\mathcal{M}}\bigcup\limits_{s=1}^{n'_{\gamma
                    l}}\xi_{\gamma l}^s\,.
\end{equation*}
The matrices ${P'}_{\gamma l}^{\,s}\in \mathfrak{Q}'_l$ are
one-dimensional projectors, pairwise orthogonal for each $\gamma$
and such that $\vee\{{P'}_{\gamma l}^s\,|\,\,k=1,...\,,n_{\gamma
l} ;\,\,\gamma\in\Sigma\}=\mathfrak{Q}'_l$ holds.
\end{theorem}
We will refer to the representation (\ref{geom form}) as
\textit{geometric form} of the eikonal algebra.

\subsubsection*{Frame ${\mathfrak F}^{\,\rm g}_\Sigma$}
Further considerations deal with the optimal partition (\ref{new
Phi++}) and the corresponding geometric form (\ref{geom form}).
Simplifying the notation, we remove the primes: $\Phi'_l=:\Phi_l$,
$\Lambda'_l=:\Lambda_l$, etc.
\smallskip

\noindent$\bullet\,\,\,$ Each point $x=x(r)\in \Phi_l$ is an
element of its minimal set $\Lambda_l[x(r)] =
\{x_1^l(r),\dots,x_{m_l}^l(r)\}$. Each family $\Phi_l$ possesses
the {\it boundaries}
\begin{equation*}
\Lambda_l^{-} := \lim\limits_{r\to 0}\Lambda_l[x(r)],\quad
\Lambda_l^{+} := \lim\limits_{r\to \epsilon_{l}}\Lambda_l[x(r)].
\end{equation*}
From (\ref{new Phi++}) we have:
\begin{equation}\label{crit points base}
\Theta = \bigcup_{l=1}^{\mathcal{M}}\left[\Lambda_l^{-}\cup
\Lambda_l^{+}\right].
\end{equation}
On the set of all boundaries $\{
\Lambda=\Lambda_l^{\pm}\,|\,\,l=1,\dots,\mathcal{M}\}$ we
introduce the relation $\sim_0$: $\Lambda\sim_0\Lambda'$ if
$\Lambda\cap \Lambda' \neq \emptyset$. Let $\sim$ be the
transitive closure of $\sim_0$ and $[\Lambda]$ be the equivalence
class of ${\Lambda}$. The representation (\ref{crit points
base}) is transformed to a partition:
\begin{equation*}
\Theta= [\Lambda]_1\cup\dots\cup[\Lambda]_K.
\end{equation*}
The set of classes
$$
\mathscr W^{\rm g}\,:=\,\left\{w_1,\dots,w_K\right\},\quad
w_k:=[\Lambda]_k
$$
will play the role of the set of {\it vertices} of the frame to be
constructed. Let us mention that there are also possible vertices
of valence 2.
\smallskip

\noindent$\bullet\,\,\,$ Let us say that a family $\Phi_l$ is
adjacent to a vertex $w=[\Lambda]$ if at least one of its
boundaries $\Lambda^\pm_l$ lies in $[\Lambda]$. By identifying the
points ${x}^l_1(r),\dots,{x}^l_{m_l}(r)$ and making up the set
$\Lambda[x(r)]\subset\Phi_l$, we turn the family into an {\it
edge}
$$
\lambda_l:=\{\lambda_l(r)\,|\,\,0<r<\epsilon_l\},\quad
\lambda_l(r):={x}^l_1(r)\equiv\dots\equiv{x}^l_{m_l}(r) ,
$$
adjacent to the vertex $w$. Each vertex has its own set of
adjacent edges $\lambda_{i_1}\dots,\lambda_{i_{d_w}}$; the number
$d_w$ is its {\it valency}. By $\mathscr E^{\rm
g}:=\{\lambda_1,\dots,\lambda_\mathcal M\}$ we denote the set of
edges.

The edges are metricized: $ {\rm
dist\,}(\lambda_l(r),\lambda_l(r')):=|r-r'| $, and equipped with
$\gamma$-coordinates:
\begin{equation}\label{coord points+}
\sigma_{\gamma}(\lambda(r)):= \{\tau_{\gamma
l}^1(r),\dots,\tau_{\gamma l}^{n_{\gamma l}}(r)\},\quad
\gamma\in\Sigma
\end{equation}
(see (\ref{geom form})).
\smallskip

\noindent$\bullet\,\,\,$ The set ${\mathfrak F}^{\,\rm
g}_\Sigma:=\mathscr W^{\rm g}\sqcup\mathscr E^{\rm g}$ is called
the \textit{geometric frame} of ${\Omega^T[\Sigma]}$.

Using the same trick (\ref{metric tau}) that equips stars with the
metric, the metric from the edges extends to the whole frame,
turning it into a metric graph.

The edges are equipped with coordinates (\ref{coord points+}); for
the vertices we put

\begin{equation*}
 \sigma_\gamma(w):= \bigcup_{\lambda_l\,\text{adjacent to $w$}} \lim\limits_{\lambda_l(r)\to
w}\sigma_{\gamma}(\lambda(r)),\quad
\gamma\in\Sigma
\end{equation*}
(the union of numerical sets on a common numerical axis).

As a result, the frame ${\mathfrak F}^{\,\rm g}_\Sigma$ is endowed
with coordinates:
\begin{equation}\label{sigma gamma for frame spec+}
{\mathfrak F}^{\,\rm
g}_\Sigma\ni\lambda\mapsto\{\sigma_\gamma(\lambda)\,|\,\,\gamma\in\Sigma\}.
\end{equation}

\noindent$\bullet\,\,\,$ In quite the same way as in the
construction of the functional model on the frame ${\mathfrak
F}^{\,\rm a}_\Sigma$, the matrix-valued functions which constite
the algebra $U\mathfrak{E}_{\Sigma}
U^{-1}\subset\bigoplus\limits_{l=1}^\mathcal M
C\left([0,\epsilon_l] \mathfrak{Q}_l\right)$ (see (\ref{geom
form})), can be transferred to the frame ${\mathfrak F}^{\,\rm
g}_\Sigma$ and thus one obtains the second functional model of
eikonal algebra. It corresponds to the optimal partition of the
domain $\Omega^T[\Sigma]$.

\section{Ordinary graphs}

\subsubsection*{Identity of frames}
\noindent$\bullet\,\,\,$ Searching the passage from the source
parametric form (\ref{param form basic}) to the canonical forms,
it is easy to recognize that the possible difference between them
is due to the difference in the partitions (\ref{P irreduce}) and
(\ref{P irreduce+}). The definitions follow to
$[P]^j_q\subset\langle P\rangle^j_q$, so the difference is
possible if and only if at least one of the classes $\langle
P\rangle^j_q$ in (\ref{P irreduce+}) admits a non-trivial
decomposition with respect to $\overset{\rm nort}\sim$.
\begin{definition}
We say that the domain $\Omega^T[\Sigma]$ is an { \it ordinary
graph} if the relations $\overset{\rm nort}\sim$ and $\overset{\rm
supp}\sim$ are equivalent on all sets $\mathbb P^j,
\,\,\,j=1,\dots,J$, what is equivalent to the identity of
decompositions (\ref{P irreduce}) and (\ref{P irreduce+})
(identities of classes $[P]^j_q=\langle P\rangle^j_q$ for all
$q=1,\dots,p_j=q_j$).
\end{definition}
It seems that a special "tuning"\ of the graph $\Omega^T[\Sigma]$
parameters {(lengths of edges)} is required to break
ordinariness; in particular, a proper choice of the value $T$.
Therefore, it is probably reasonable to speak of ordinariness as a
generic case.
\smallskip

The following statement is valid:
\begin{theorem}
Let the domain $\Omega^T[\Sigma]$ be an ordinary graph. Then the
correspondence between the frames ${\mathfrak F}^{\,\rm
a}_\Sigma\ni
{\mbox{\boldmath$\pi$}}{\leftrightarrow\lambda\in{\mathfrak
F}^{\,\rm g}_\Sigma}$ defined by the relation
\begin{equation}\label{equiv coord}
\scalemath{1.35}{\mbox{\boldmath$\sigma$}}_{\gamma}(\scalemath{1.35}{\mbox{\boldmath$\pi$}})\,=\,\sigma_\gamma(\lambda),\qquad\gamma\in\Sigma
\end{equation}
turns out to be {\it isometry} of the frames (as metric spaces).
\end{theorem}

\noindent{\bf Proof}\,\,\,Ordinariness is equivalent  to
identity of classes in (\ref{P irreduce}) and (\ref{P irreduce+}):
$[P]^j_q=\langle P\rangle^j_q$. The identity leads to equality of
algebras $\mathfrak P_l=\mathfrak Q_l$ and their simultaneous
irreducibility. Thus, the representations (\ref{param form step
1}) and (\ref{param form step 1+}) are identical.

Using the representations (\ref{param form step 1}) and
(\ref{param form step 1+}), we can construct isometric copies of
frames ${\mathfrak F}^{\,\rm a}_\Sigma$ and ${\mathfrak F}^{\,\rm
g}_\Sigma$. Let us describe this procedure for ${\mathfrak
F}^{\,\rm a}_\Sigma$; for ${\mathfrak F}^{\,\rm g}_\Sigma$ all
considerations are similar.

With each block $[U\mathfrak{E}_{\Sigma}U^{-1}]_l $ we associate an open interval
$${\omega_l}:={\{x(r):=r\,|\,\,r\in (0,\epsilon_l)\}\subset \mathbb{R}} $$
equipped with $\gamma-$coordinates
$$
{ \sigma(x(r)):=\{\tau_{\gamma
l}^k(r)|\,\,\gamma,l,k\,\,\,\text{are such that}\,\,\, P_{\gamma
l}^k \in[P]_l\,\,\,\text{holds}\}.}
$$ Let us also consider its closure $\overline{\omega_l}$, where
the coordinates of two endpoints $q^{-}_l,q^{+}_l\in
\overline{\omega_l}\setminus \omega_l$ are defined by continuity.
Consider the set of all interval endpoints $\Upsilon:=
\{q_l^{-},q_l^{+}|\,\,l=1,\dots, M\}$ and introduce the relation
$\sim_0$ on it by the rule: $q_l^c \sim_0 q_{l'}^{c'}$ if
$\sigma_{\gamma}(q_l^c)\cap \sigma_{\gamma}(q_{l'}^{c'})\neq
\emptyset$ for some $\gamma\in\Sigma$, {where $c,c'\in
\{-,+\}$}. By $\sim$ we denote the transitive closure of the
relation $\sim_0$. Then the set $\Upsilon$ can be represented as a
disjunct union of equivalence classes $[q]_k$ under this relation:
    $$
    \Upsilon = [q]_1\cup\dots\cup [q]_K.
    $$
Consider $w_k:=[q]_k$; the elements of the set $W:=\{
w_1,\dots,w_K\}$ will be called vertices, the elements of the set
$E:=\{ \omega_1,\dots,\omega_M\}$ -- edges, and the set
$\tilde{\mathfrak{ F}}^{\rm a}_{\Sigma}  := W \sqcup E$ -- an
auxiliary algebraic frame. For the points of the edges, the
$\gamma-$coordinates were defined above, and for the vertices we
put
    $$
    \sigma_{\gamma}(w):=\bigcup\limits_{q\in w} \sigma_{\gamma}(q).
    $$
Note that the frame $\tilde{\mathfrak{ F}}^{\rm a}_{\Sigma}$ may
differ from $\mathfrak{ F}^{\rm a}_{\Sigma}$ only by the presence
of additional valency 2 vertices.

The auxiliary geometric frame $\tilde{\mathfrak{ F}}^{\,\rm
g}_\Sigma $ is constructed quite analogously. It is easy to see
that if $\Omega^T[\Sigma]$ is ordinary, then the auxiliary frames
are identical: $\tilde{\mathfrak{ F}}^{\,\rm a}_\Sigma \equiv
\tilde{\mathfrak{ F}}^{\,\rm g}_\Sigma$.

In the mean time, as is easy to see, each of the auxiliary frames
is isometric to the original one (algebraic or geometric), whereas
the isometries are set up by bijections defined by equality of
coordinates (\ref{equiv coord}). Thus, the ordinariness of the
domain $\Omega^T[\Sigma]$ does lead to isometry between the {
algebraic and geometric} frames ${\mathfrak{ F}}^{\,\rm
a}_\Sigma  $ and $ {\mathfrak{ F}}^{\,\rm g}_\Sigma $.\qquad\qquad\qquad\qquad\qquad\qquad\qquad\qquad\qquad\qquad\qquad\qquad$\square$

\subsubsection*{Commentary}

\noindent$\bullet$\,\,\,Using standard techniques of the Boundary
Control method \cite{Bel_UMN}, it can be shown that the
traditional inverse problem data define some {\it isomorphic copy}
$\left[\mathfrak{E}_{\Sigma}\right]^c$ of the algebra
$\mathfrak{E}_{\Sigma}$. As a consequence, these data determine an
{\it isometric copy} $\left[{\mathfrak F}^{\,\rm
a}_\Sigma\right]^c$ of the frame ${\mathfrak F}^{\,\rm a}_\Sigma$.
If the domain $\Omega^T[\Sigma]$ filled with waves is an ordinary
graph, then we have an isometric copy $\left[{\mathfrak F}^{\,\rm
g}_\Sigma\right]^c$ of the frame ${\mathfrak F}^{\,\rm g}_\Sigma$.
Just like this frame itself, its copy corresponds to the optimal
partition and, hence, contains information about the graph
$\Omega^T[\Sigma]$ geometry. We can try to obtain this information
by the following scheme:
$$
\text{Inverse problem data}\,
\Rightarrow\,\left[\mathfrak{E}_{\Sigma}\right]^c\,
\Rightarrow\,\left[{\mathfrak F}^{\,\rm
a}_\Sigma\right]^c\equiv\left[{\mathfrak F}^{\,\rm
g}_\Sigma\right]^c\,\overset{?}\Rightarrow\,\Omega^T[\Sigma]
$$
The fundamental question that sets the direction for further
investigation is to what extent the copy $\left[{\mathfrak
F}^{\,\rm g}_\Sigma\right]^c$ determines this geometry. To put it
in simpler terms, can one "unglue"\, ${\mathfrak F}^{\,\rm
g}_\Sigma$ into $\Omega^T[\Sigma]$? The question is open and seems
to be rather complicated.
\smallskip

\noindent$\bullet$\,\,\,The question on the validity of the
following hypothesis stated in \cite{Belishev_Wada_2015} also
remains open. In the known examples, each vertex $v$ covered
by the waves from (at least) two controlling vertices, i.e. such
that $v\in\Omega^T[\gamma]\cap\,\Omega^T[\gamma']$ is satisfied,
corresponds to a cluster in the spectrum $\widehat{\mathfrak
E_\Sigma}$. Is it always valid? Also, is it possible to detect
the presence or absence of cycles in $\Omega^T[\Sigma]$ from the
spectrum $\widehat{\mathfrak E_\Sigma}$ (frame ${\mathfrak
F}^{\,\rm g}_\Sigma$)\,\, \cite{Belishev_Wada_2009}\,?

\end{document}